\documentclass[12pt,preprint]{aastex}
\usepackage{graphicx}

\begin{document}

\def\lesssim{\mathrel{\hbox{\rlap{\hbox{\lower4pt\hbox{$\sim$}}}\hbox{$<$}}}}

\def\gtrsim{\mathrel{\hbox{\rlap{\hbox{\lower4pt\hbox{$\sim$}}}\hbox{$>$}}}}



\title{Asteroseismological study of massive ZZ Ceti stars with 
fully evolutionary models}

\author{A. D. Romero$^{1}$,
S. O. Kepler$^{1}$,
A. H. C\'orsico$^{2,3}$, 
L. G. Althaus$^{2,3}$, and
L. Fraga$^{4,5}$}

\affil{ $^{1}$Departamento de Astronomia, Universidade Federal do 
Rio Grande do Sul, Av. Bento Goncalves 9500
Porto Alegre 91501-970, RS, Brazil\\
$^{2}$Facultad de Ciencias Astron\'omicas y Geof\'{i}sicas, 
Universidad Nacional de La Plata, Paseo del Bosque s/n, (1900) La Plata, Argentina\\
$^{3}$Instituto de Astrof\'{i}sica La Plata, CONICET-UNLP, Argentina\\
$^{4}$Southern Observatory for Astrophysical Research, Casilla 603, La Serena, Chile\\
$^{5}$Laborat\'{o}rio Nacional de Astrof\'{i}sica -- LNA/MCTI, R. Estados Unidos, 154, Itajub\'{a}, MG, CEP: 37504-364, Brazil}

\email{alejandra.romero@ufrgs.br}

\begin{abstract}

We present the first asteroseismological  study for 42 massive ZZ Ceti
stars based on a large  set of fully evolutionary carbon$-$oxygen core
DA  white dwarf  models  characterized by  a  detailed and  consistent
chemical  inner profile  for the  core  and the  envelope. Our  sample
comprise  all the  ZZ  Ceti stars  with  spectroscopic stellar  masses
between   $0.72$    and   $1.05M_{\odot}$   known    to   date.    The
asteroseismological  analysis   of  a  set  of  42   stars  gives  the
possibility to study the  ensemble properties of the massive pulsating
white  dwarf  stars  with  carbon$-$oxygen cores,  in  particular  the
thickness  of   the  hydrogen  envelope  and  the   stellar  mass.   A
significant fraction  of stars  in our sample  have stellar  mass high
enough as to crystallize at  the effective temperatures of the ZZ Ceti
instability  strip,   which  enables  us  to  study   the  effects  of
crystallization  on  the pulsation  properties  of  these stars.   Our
results  show that  the phase  diagram  presented in  Horowitz et  al.
(2010)  seems  to be  a  good  representation  of the  crystallization
process inside white  dwarf stars, in agreement with  the results from
white  dwarf  luminosity  function  in globular  clusters.

\end{abstract}



\section{Introduction}

ZZ  Ceti (or  DAV) stars  are the  most numerous  class  of degenerate
pulsators,  with $\sim  160$  members known  to  date (Castanheira  et
al. 2013a).  These stars have  hydrogen atmospheres and are located in
a narrow range in effective temperature $10\, 500 \lesssim T_{\rm eff}
\lesssim 12\, 300$ K (e.g. Fontaine \& Brassard 2008; Winget \& Kepler
2008; Althaus  et al.  2010a), mostly with  temperatures close  to the
center  of the  instability  strip, and  masses  $\sim 0.6  M_{\odot}$
(Gianninas, Bergeron  \& Fontaine  2005: Castanheira \&  Kepler 2008).
Their  photometric  variations   are  due  to  spheroidal,  non-radial
$g$-mode pulsations  with  low  harmonic  degree ($\ell  \leq  2$)  and
periods in the range 70$-$2000 s, with amplitude variations of up to
0.3 mag.  The  driving mechanism thought to excite  the pulsation near
the  blue edge  of  the instability  strip  is the  $\kappa -  \gamma$
mechanism  acting on the  hydrogen partial  ionization zone  (Dolez \&
Vauclair 1981; Winget et al.  1982).  The
``convective  driving mechanism'' proposed  first by  Brickhill (1991)
and later  revisited by  Goldreich \& Wu  (1999) is thought  to become
dominant  once a  thick convective  zone  has developed  in the  outer
layers.

The ZZ  Cetis can  be classified into  three groups, depending  on the
effective temperature (Mukadam et al.  2006, Clemens et al. 1993). The
hot ZZ Cetis, that define the  blue edge of the instability strip, and
exhibit  a  few  modes  with  short  periods ($<  350$  s)  and  small
amplitudes  (1.5-20 mma). The  pulse shape  is sinusoidal  or sawtooth
shaped  and  is stable  for  decades.  On  the  opposite  side of  the
instability strip  we have  the cool DAV  stars, showing  several long
periods (up  to 1500 s), with  large amplitudes (40-110  mma), and non
sinusoidal light  curves that can  change dramatically from  season to
season due  to the mode  interference. Finally, Mukadam et  al. (2006)
suggested introducing  a third class, the intermediate  ZZ Cetis, that
exhibit mixed characteristics from  hot and cool DAV stars.  Recently,
Hermes  et  al.  (2012,  2013a)  extended  the  variability  strip  of
pulsating DAV stars  to cooler temperatures with the  discovery of low
mass  pulsators.   The  variable   low  mass  white  dwarf  stars  are
characterized by  effective temperatures below  $\sim 10\, 000$  K and
long periods in the range $1000-4500$ s.

Over  the   years,  pulsation  studies   of  ZZ  Ceti   stars  through
asteroseismology have become a valuable technique to study the details
of  the  origin, internal  structure  and  evolution  of white  dwarfs
(Winget  \&  Kepler  2008;  Fontaine  \&  Brassard  2008,  Althaus  et
al. 2010a).  In  particular, the thickness of the  outer envelope, the
chemical composition  of the core, magnetic fields  and rotation rates
can be determined from the  observed periods.  Also the rate of period
change  can be  employed  to  measure their  cooling  rate (Kepler  et
al. 2005) and to study  particles like neutrinos (Winget et al.  2004)
or axions (Isern et al.   1992: C\'orsico et al. 2001; Bischoff-Kim et
al. 2008; Isern et al 2010; C\'orsico et al.  2012ab) and the possible
rate of variation of the  Newton constant (C\'orsico et al. 2013), but
also to look  for extrasolar planet orbiting these  stars (Mullally et
al.  2008).  In  addition, asteroseismology of white dwarf  stars is a
valuable tool to place observational constraints on the crystallization
process in the  very dense interiors of white  dwarf stars (Montgomery
\& Winget  1999; C\'orsico  et al. 2004,  2005; Metcalfe et  al. 2004;
Kanaan et al. 2005).

The number  of white dwarf stars,  and consequently of  ZZ Ceti stars,
has dramatically  increased with the  Sloan Digital Sky  Survey (SDSS)
project  (Mukadam et  al.   2004;  Mullally et  al.   2005; Kepler  et
al.  2005, 2012;  Castanheira et  al.  2006,  2007, 2010,  2013a).  In
particular, Kleinman et  al. (2013) reported $12\, 843$  DA and 923 DB
white  dwarfs  stars  from  the  SDSS Data  Release  7  (Abazajian  et
al. 2009). The mass distribution for DA white dwarf stars presented by
Kleinman et al. (2013) shows a main components with stellar mass
  around $\sim 0.59 M_{\odot}$, that comprise $\sim 80\%$ of the total
  sample,  and also a  low$-$mass and  a high$-$mass  components. The
main  population   of  high$-$mass  white  dwarfs   has  masses  above
0.721$M_{\odot}$ and  peaks at 0.822 $M_{\odot}$.  It  is thought that
most of the  white dwarf stars populating the  high-mass component are
likely to have carbon$-$oxygen  cores, formed during the stable helium
burning   phase  in   the  pre$-$white   dwarf   evolution.   However,
evolutionary  computations show that  stars with  stellar mass  in the
Zero Age  Main Sequence (ZAMS)  of $\gtrsim 7 M_{\odot}$  reach stable
carbon   ignition   giving   rise    to   an   oxygen$-$neon   or   an
oxygen$-$neon$-$magnesium  core  white dwarf  star  with stellar  mass
larger than $\sim  1.05 M_{\odot}$ (Ritossa et  al.  1999; Siess
2007).  Therefore,  the massive component  of the DA white  dwarf mass
distribution  is  populated  by  carbon$-$oxygen core  stars,  with  a
progenitor star of $\sim 3-7 M_{\odot}$, as well as oxygen$-$neon core
white  dwarf stars  with  stellar mass  above  $\sim 1.05  M_{\odot}$,
resulting  from a  progenitor  star with  masses  between $\sim  7-8.5
M_{\odot}$ \footnote{ The  upper limit for the progenitor  star is set
  by observations  of Type II  supernova (Smartt 2009),  although this
  value should be metallicity dependent.}.

High$-$mass white dwarf  stars are not easy to  find, not only because
their intrinsically  smaller number with  respect to lower  mass white
dwarfs, but also because they  evolve fast and have low luminosity due
to  their   smaller  radius.   Therefore,  the   search  for  variable
high$-$mass  white  dwarfs is  quite  challenging.   The most  studied
massive DAV  star is BPM  37093 with $M_*  = 1.1 \pm  0.05 M_{\odot}$,
discovered  to be variable  by Kanaan  et al.  (1992).  Because  of its
high mass,  BPM  37093 was  considered  the  only  pulsator to  have
undergone partial  crystallization and thus  presented the opportunity
to study the crystallization theory through asteroseismology (Metcalfe
et al.  2004; Kanaan et al.  2005).  In the past  few years, pulsation
variability  have  been searched  for  and  detected  in several  other
massive  DAV stars  (Kepler  et  al. 2005;  Castanheira  et al.  2006;
Castanheira  et   al.  2010;  Kepler  et  al.   2012,  Castanheira  et
al. 2013a), but  only a few of these objects  have been analyzed using
asteroseismology.  The  first attempt to study  the general properties
of massive DAV stars has been presented recently by Castanheira et al.
(2013a).   In addition  to report  the discovery  of five  new massive
pulsators and perform seismological fits for these particular objects,
they carried out  a study of the observational properties  of a set of
massive pulsating DA stars with spectroscopic stellar mass higher than
$0.8  M_{\odot}$.  They  choose only  25 stars  with SDSS  spectra in
order  to   have  an  homogeneous  sample  in   terms  of  atmospheric
determinations (Kleinman et al. 2004, Kleinman et al. 2013).

Currently,  there is  no asteroseismological  study in  the literature
applied specifically to the massive  variable white dwarfs as a group.
This paper is  intended to fill this gap.   Specifically we perform an
asteroseismological analysis  of all  massive DA variable  white dwarf
stars known to date, with spectroscopic masses in the range $0.72-1.05
M_{\odot}$.  Note that the stellar mass values in the target selection
include only  massive DAVs expected to have  carbon$-$oxygen cores. We
defer the study  of very massive DAVs thought  to harbor oxygen$-$neon
core DAVs, like the classical  BPM 37093 (Kanaan et al. 1998; Metcalfe
et  al.  2004)  and  the recently  discovered  GD 518  (Hermes et  al.
2013b), for a future paper  as our current evolutionary models are not
appropriate.   Our sample  of  massive  DAV stars  is  composed by  42
objects, 36 of which were  discovered within the SDSS DR7 (Kleinman et
al.   2013), 5  objects are  bright ZZ  Ceti stars  (e.g.  Fontaine \&
Brassard 2008), and  one was selected from the  SuperCOSMOS Sky Survey
catalog  (Rowell   \&  Hambly  2011).    Most  of  these   stars  have
observational data only from the discovery paper with only a few modes
detected, in many  cases a single mode.  To  perform our seismological
study we employ  a grid of full evolutionary  models representative of
white  dwarf  stars discussed  in  Romero  et  al. (2012)  which  have
consistent chemical  profiles for both  the core and the  envelope for
various   stellar   masses,   particularly   intended   for   detailed
asteroseismological fits  of ZZ Ceti stars.  The  chemical profiles of
our models  are computed from the  full and complete  evolution of the
progenitor  stars from  the ZAMS,  through the  thermally  pulsing and
mass$-$loss  phases  on  the   asymptotic  giant  branch  (AGB). We 
consider the ocurrence of extra$-$mixing episodes during all stages prior 
to the thermally pulsing AGB, and time$-$dependent element diffusion  during the white dwarf
stage   (Althaus  et   al.   2010b,  Renedo   et   al.   2010).    Our
asteroseismological approach combines  $(i)$ a significant exploration
of the  parameter space $(M_*, T_{\rm  eff}, M_{\rm H}$)  and $(ii)$ a
detailed  and  updated input  physics,  in  particular, regarding  the
internal structure, that is a crucial aspect for correctly disentangle
the information  encoded in the  pulsation patterns of  variable white
dwarfs.  The first  version of this model grid  was employed by Romero
et al.  (2012) to perform  an asteroseismological study of a sample of
44  bright   ZZ  Ceti  stars  with  stellar   masses  $\simeq  0.5-0.8
M_{\odot}$, including  G117$-$B15A. In this paper we  present a second
version  of the  model grid,  where  we extended  our parameter  space
towards higher stellar mass  values. In addition, we include different
treatments of  the crystallization process in our  computations, as an
extra parameter in our model grid.
  
The  paper is  organized  as follows.   In  Section \ref{section2}  we
describe the  evolutionary code and  the input physics adopted  in our
computations  and   we  present  the   model  grid  employed   in  our
asteroseismological study. Section \ref{crystallization} is devoted to
study the effects of crystallization process on the pulsation spectrum
of massive white dwarf stars. In Section \ref{spectroscopy} we present
our  sample  of massive  DA  variable  white  dwarfs and  quote  their
properties  from  spectroscopy.   Section  \ref{observations}  briefly
introduces  new observations  performed  for a  few  objects from  our
sample.   In  Section  \ref{fits}  we  present our  results  from  the
asteroseismological    procedure.      We    conclude    in    Section
\ref{conclusions} by summarizing our findings.
 
\section{Numerical tools and models}
\label{section2}

\subsection{Input physics}
\label{input physics}

The  grid of  full  evolutionary  models employed  in  this work  were
calculated  with an updated  version of  the LPCODE  evolutionary code
(see Althaus et al.  2005a; Althaus et al.  2010b; Renedo et al.  2010
for details).   Here, we comment  the main input physics  relevant for
this work. Further details can be found in those papers.

The  LPCODE evolutionary  code considers  a simultaneous  treatment of
no$-$instantaneous  mixing   and  burning  of   elements  (Althaus  et
al. 2003). The nuclear network accounts explicitly for 16 elements and
34 nuclear reactions, that include $pp-$chain, CNO$-$cycle, helium burning
and  carbon  ignition  (Renedo  et  al.  2010).   In  particular,  the
$^{12}$C$(\alpha  ,  \gamma  )   ^{16}$O  reaction  rate,  of  special
relevance  for  the carbon$-$oxygen  stratification  of the  resulting
white dwarf, was  taken from Angulo et al.  (1999).  As a consequence,
our  white  dwarf models  are  characterized  by systematically  lower
central  oxygen  abundances  than  the  values  found  by  Salaris  et
al. (1997),  who use the  larger rate of  Caughlan et al.   (1985) for
this  reaction. 

Also,  we consider  the occurrence  of extra$-$mixing episodes beyond
each convective  boundary following the prescription of  Herwig et al.
(1997), except for the  thermally pulsating AGB phase.  The occurrence
of  extra$-$mixing  episodes  during  the core  helium  burning  phase
largely  determines  the  final   core  chemical  composition  of  the
resulting white  dwarf (Straniero et  al.  2003). We treated the extra$-$mixing (overshooting) as a time dependent diffusion process --- by assuming that the mixing velocities decay exponentially beyond each convective boundary --- with a diffusion coefficient given by $D_{\rm EM}=D_0 \exp(-2z/f H_P)$, where $H_P$ is the pressure scale height at the convective boundary, $D_O$ is the diffusion coefficient of unstable regions close to the convective boundary, and $z$ is the geometric distance from the edge of the convective boundary (Herwing et al. 1997, 2000). The free parameter $f$ describes the efficiency of the extra$-$mixing process. It can take values as high as $f\sim 1.0$, for overadiabatic convective envelopes of DA white dwarfs (Freytag et. al 1996). However, for deep  envelope and  core convection  $f$ is  expected  to be considerably  smaller because  the ratio  of  the Brunt$-$V\"ais\"al\"a timescales of the stable to  unstable layers decrease with depth. In this study we have assumed f = 0.016, which accounts for the location of the upper envelope on the main sequence for a large sample of clusters and associations (Schaller et al. 1992, Herwig et al. 1997, 2000). Also it accounts for the intershell abundances of hydrogen$-$deficient post$-$AGB remnants (see Herwig et al. 1997; Herwig 2000; Mazzitelli et al. 1999). Finally, for the mass range considered in this work the mass  of  the outer  convection  zone on the tip of the RGB  only increases by $\sim1.5 \%$ if  we change the parameter $f$ by a factor of two. 
The  suppression of
extra$-$mixing  events  during   the  thermally  pulsating  AGB  phase
prevents the third  dredge$-$up to occur in low  mass stars (Lugaro et
al. 2003; Herwig  et al.  2007; Salaris et al.   2009), leading to the
gradual increase of the hydrogen  free core mass as evolution proceeds
during this phase.  As a result, the initial$-$final mass relationship
by the end  of the thermally pulsating AGB  is markedly different from
that  resulting from  considering the  hydrogen free  core  mass right
before  the first  thermal pulse.   In fact,  Althaus et  al.  (2010b)
demonstrated that, depending  on the stellar mass of  the white dwarf,
the central  oxygen abundances can be underestimated  up to a $\sim$15\% if
the white dwarf mass is taken as the mass of the hydrogen free core at
the  first thermal  pulse.  Finally,  the breathing  pulse instability
occurring  towards the  end of  the  core helium  burning are  usually
attributed  to the  adopted algorithm  rather than  to the  physics of
convection  and therefore  were  suppressed in  our computations  (see
Straniero et al.  2003 for a more detailed discussion).

We considered  mass loss  during the core  helium burning and  the red
giant branch  phases following Schr\"oder \& Cuntz (2005),  and during
the  AGB and  thermally pulsating  AGB following  the  prescription of
Vasiliadis \&  Wood (1993). Since there  is a strong  reduction of the
third dredge$-$up,  as is the case  of the sequences  computed in this
work, mass loss plays an  important role in determining the final mass
of the  hydrogen free core at  the end of the  thermally pulsating AGB
evolution, and  thus the initial$-$final mass  relation.  However, the
residual  helium burning  in a  shell,  that increases  the core  mass
during the  thermally pulsating  AGB and the  hot stages of  the white
dwarf evolution, is also important in determining the white dwarf  final 
mass.  

During the  white dwarf evolution,  we consider the  distinct physical
process  that   modify  the  chemical   abundances  distribution.   In
particular,   element   diffusion   strongly  affects   the   chemical
composition profile  throughout the  outer layers. Indeed, our
sequences developed a pure hydrogen envelope with increasing thickness
as evolution proceeds.  Our treatment of time$-$dependent diffusion is
based on the multicomponent gas treatment presented in Burgers (1969).
We consider gravitational settling  and thermal and chemical diffusion
of  $¹$H, $^3$He,  $^4$He, $^{12}$C,  $^{13}$C, $^{14}$N  and $^{16}$O
(Althaus et al.  2003).  To  account for convection process we adopted
the mixing length  theory, in its ML2 flavor,  with the free parameter
$\alpha =  1.61$ (Tassoul  et al.  1990).  Finally, we  considered the
chemical  rehomogenization  of   the  inner  carbon$-$oxygen  profile
induced  by Rayleigh$-$Taylor instabilities  following Salaris  et al.
(1997).

The  input  physics of  the  code include  the  equation  of state  of
Segretain et al. (1994) for  the high density regime complemented with
an updated  version of  the equation of  state of Magni  \& Mazzitelli
(1979)  for  the  low  density  regime.   Other  physical  ingredients
considered in LPCODE are the radiative opacities from the OPAL opacity
project  (Iglesias \&  Rogers 1996)  supplemented at  low temperatures
with the  molecular opacities of Alexander \&  Ferguson (1994). During
the  white dwarf  cooling, the  metal mass  fraction $Z$  is specified
consistently    according    to    the   predictions    of    chemical
diffusion. Conductive opacities are  those from Cassisi et al. (2007),
and the neutrino emission rates are  taken from Itoh et al. (1996) and
Haft et al. (1994). It is worth mentioning that the presence of rotation during the pre$-$white dwarf evolution may affect the chemical structure and the evolutionary properties of the models. For instance Georgy et al. (2013), based on a large grid of rotating and non$-$rotating MS models, found that rotation induces an increase in the MS lifetimes of about 20$-$25\% for models with initial mass larger than 1.7 $M_{\odot}$. Also there is a nitrogen enrichment at the end of the MS for models with rotation. On the other hand, the final mass for rotating models is similar to that for non$-$rotating models.

Cool  white dwarf stars  are expected  to crystallize  as a  result of
strong  Coulomb interactions in  their very  dense interior  (van Horn
1968). Crystallization occur when the energy of the Coulomb interaction between neighboring ions is much larger than their thermal energy. This occurs when the ion coupling constant $\Gamma \equiv  \langle Z^{5/3}\rangle  e^2/a_{\rm  e}k_{\rm  B}T$ is larger than certain value, which depends on the adopted phase diagram. Here $a_{\rm e}$ is the interelectronic distance, $\langle Z^{5/3}\rangle$  an average (by number) over the ion charges, and $k_{\rm B}$ Boltzmann's constant. The rest of the symbols have their usual meaning. The occurrence of crystallization leads to two additional energy
sources: the release  of latent heat and the  release of gravitational
energy  associated  with  changes   in  the  chemical  composition  of
carbon$-$oxygen profile induced by crystallization (Garc\'{i}a$-$Berro
et al.  1988ab, Winget et  al. 2009).  In  our study, the  inclusion of
these two  additional energy sources was  done self$-$consistently and
locally coupled  to the  full set of  equations of  stellar evolution,
were the luminosity equation  is appropriately modified to account for
both the local contribution of  energy released from the core chemical
redistribution   and  the   latent  heat.    
At  each   timestep,  the
crystallization  temperature and  the change  of the  chemical profile
resulting  from phase  separation  are computed  using the  appropriate
phase  diagram.   In  particular,  the  carbon  enhanced  convectively
unstable liquid layers overlying the crystallizing core are assumed to
be instantaneously mixed, a reasonable assumption considering the long
evolutionary  timescales of white  dwarfs (Isern  et al.   1997).  The
chemical redistribution  due to  phase separation has  been considered
following  the procedure described  in Montgomery  et al.   (1999) and
Salaris et  al.  (1997).   To assess the  enhancement of oxygen  in the
crystallized core  we employed two phase  diagrams: the spindle$-$type
phase diagram of Segretain \& Chabrier (1993) and the azeotropic$-$type
phase diagram of Horowitz et al.  (2010) (see Althaus et al.  2012 for
details  on   the  implementation).   
 In our computations, crystallization begins for $\Gamma \sim 180$ when we employ the Segretain \& Chabrier (1993) phase diagram and for $\Gamma \sim 220$ when we consider the Horowitz et al. (2010) phase diagram. For pure carbon crystallization  occurs when $\Gamma = 178.4$.
After   computing  the  chemical
composition of both the solid  and the liquid phases, we evaluated the
net energy  released in the  process as in  Isern et al.   (1997). 
This energy is  added to the, usually smaller, latent heat contribution, of the order of $0.77 k_B T$ per  ion. Both  energy  contributions were
distributed over a small mass range around the crystallization front.

\subsection{Model grid}
\label{model-grid}

The DA white dwarf models employed in this work are the result of full
evolutionary calculations  of the progenitor stars, from  the ZAMS, 
through the hydrogen and helium central burning stages,
thermally pulsating  and mass  loss in the  AGB phase and  finally the
planetary  nebula  domain.    They  were  generated  employing  LPCODE
evolutionary code.  The metallicity value adopted in the Main Sequence 
models is $Z  = 0.01$.  Most of  the sequences with
masses $\leq 0.878 M_{\odot}$ were employed in the asteroseismological
study of  44 bright ZZ  Ceti stars by  Romero et al. (2012),  and were
extracted  from  the  full  evolutionary computations  of  Althaus  et
al. (2010b) (see also Renedo et al. 2010).  In this work we extend the
model grid  towards the  high mass domain.  We computed five  new full
evolutionary sequences  with initial masses  on the ZAMS in  the range
$5.5 - 6.7 M_{\odot}$ resulting  in white dwarf sequences with stellar
masses between $0.917$ and $1.024 M_{\odot}$.  In addition, we compute
two  new  sequences  with  white  dwarf  masses  of  0.721  and  0.800
$M_{\odot}$  in order  to achieve  a finer  coverage of  the  low mass
region of our  sample. Finally, we obtained a  sequence with a stellar
mass 1.050  $M_{\odot}$ by artificially scaling the  stellar mass from
the   0.976$M_{\odot}$  sequence   at   high  effective   temperatures
(C\'orsico et al.  2004).  The  values of stellar mass of our complete
model grid  are listed  in column 1  of table  \ref{tabla-grid}, along
with  the  hydrogen  (column  2)  and helium  (column  3)  content  as
predicted by standard stellar  evolution, and the carbon ($X_{\rm C}$)
and oxygen ($X_{\rm  O}$) central abundances by mass  in columns 4 and
5,  respectively.  The values  of stellar  mass of  our set  of models
accounts for the  stellar mass of all the  observed pulsating DA white
dwarf  stars with a  probable carbon$-$oxygen  core.  Note  that white
dwarfs  with  masses  higher   than  1.05  $M_{\odot}$  probably  have
oxygen$-$neon cores, since they  reach off$-$center carbon ignition in
partial electron degenerate conditions before entering the white dwarf
cooling sequence.

Our parameter space  is build up by varying  three quantities: stellar
mass ($M_*$),  effective temperature ($T_{\rm eff}$)  and thickness of
the hydrogen  envelope ($M_{\rm H}$).   Both the stellar mass  and the
effective temperature  vary in a consistent way  with the expectations
from evolutionary computations. On the  other hand, we decided to vary
the  thickness  of  the  hydrogen  envelope in  order  to  expand  our
parameter space.  The choice of  varying $M_{\rm H}$ is not arbitrary,
since there  are uncertainties related  to mass loss rates  during the
AGB phase leading to uncertainties on the amount of hydrogen remaining
on the  envelope of white dwarf  stars.  For instance,  Althaus et al.
(2005b) have  found that  the amount  of hydrogen left  in a  DA white
dwarf can be significantly reduced if the progenitor star experience a
late thermal  pulse.  Tremblay  \& Bergeron (2008)  show that  a broad
range  in the  thickness  of the  hydrogen  envelope can  lead to  the
observed increase in the He$-$  to H$-$rich white dwarfs, although the
mixing  has to occur  for temperatures  lower than  $\sim 10\,  000$ K
(Tremblay et al.  2010).  Recently,  Kurtz et al.  (2013) reported the
discovery of a member of the so called ``hot DAV'' stars in the cooler
edge  of the  DB gap,  whose  pulsation instability  was predicted  by
Shibahashi (2005, 2007)  based on the possible existence  of very thin
hydrogen  envelope DA white  dwarfs ($M_{\rm  H} \sim  10^{-12} M_*$).
The remaining amount of hydrogen in  a white dwarf also depends on the
metallicity  adopted for the  progenitor star.  Renedo et  al.  (2010)
show that, for  a $\sim 0.6 M_{\odot}$, the  thickness of the hydrogen
envelope increases by a factor of $\sim 2$ when the metallicity of the
progenitor star is reduced from $Z=0.01$ to $Z=0.001$.

Other  structural parameters  are not  thought to  change considerably
according to standard evolutionary computations.  For instance, Romero
et al.   (2012) showed that the  remaining helium content  of DA white
dwarf stars cannot be substantially  lower (as much as 3$-$4 orders of
magnitude) than that predicted by standard stellar evolution, and only
at the expense of an increase in the the hydrogen free core ($\sim 0.2
M_{\odot}$).  The  structure of carbon$-$oxygen  chemical profiles are
basically fixed by the evolution  during the core helium burning stage
and are not expected to vary during the following single evolution (we
do not consider possible merger episodes)\footnote{Different metallicity values and helium contents for the progenitor stars can lead to differences on the evolution of the star. For instance, the mass of the resulting white dwarf increases in $\sim 15 \%$ when the metallicity decreases from $Z=0.01$ to $Z=0.001$ (Renedo et al. 2010). We will study the effect of these parameters in the future.}

\begin{figure}
\begin{center}
\includegraphics[clip,width=230pt]{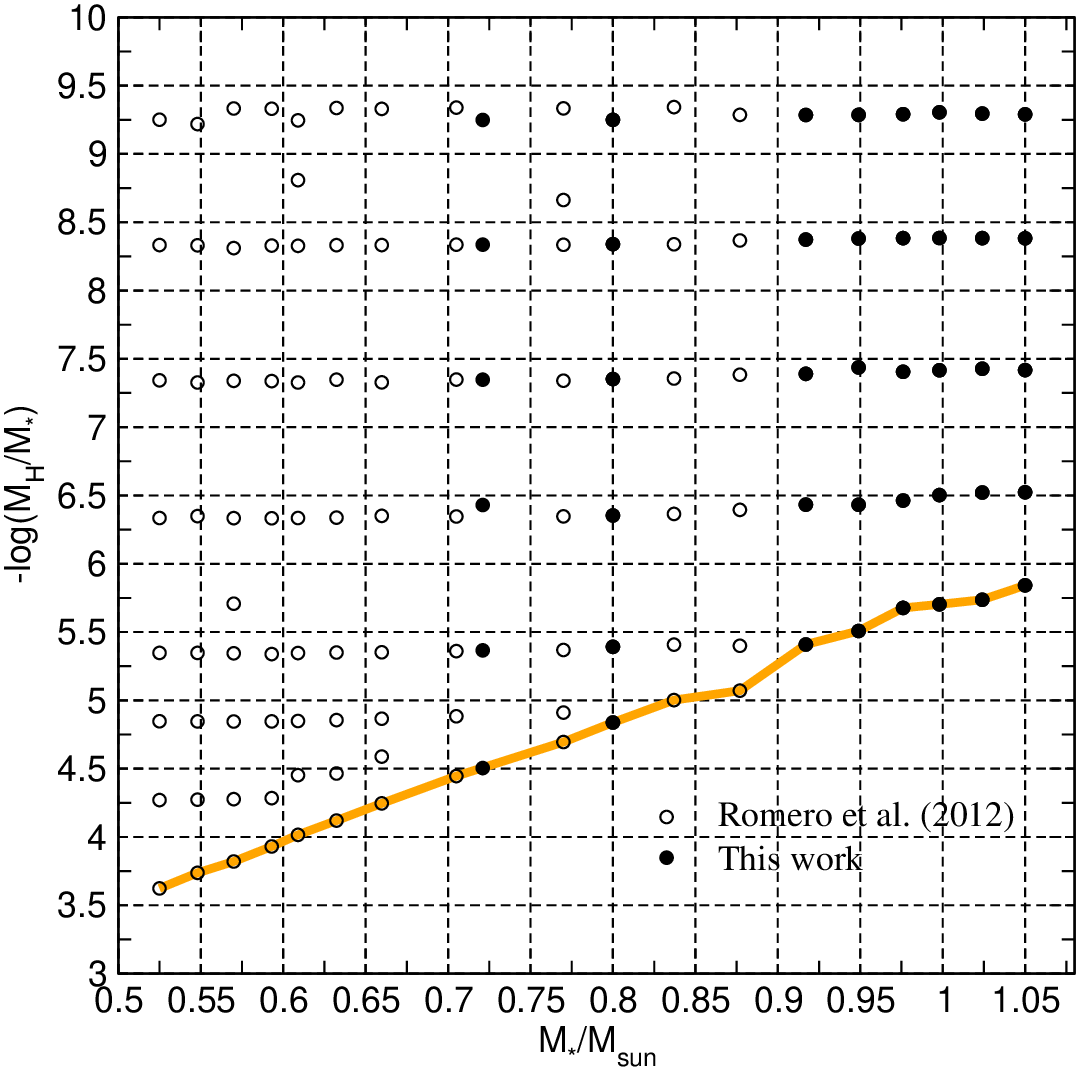}
\caption{The grid of 163 DA white dwarf evolutionary sequences 
considered in this work in the $M_* - \log(M_{\rm H}/M_*)$ plane. 
Each circle corresponds to a sequence of models representative 
of white dwarf stars characterized by a given stellar mass and hydrogen 
envelope mass. Open circles correspond to the evolutionary 
sequences computed in Romero et al. (2012), while filled circles 
correspond to sequences computed in this work. The orange thick 
line connects the sequences with the maximum values of the thickness
 of the hydrogen envelope, predicted by our evolutionary computations. }
\label{envolturas-todas}
\end{center}
\end{figure}

In  order to get  different values  of the  thickness of  the hydrogen
envelope,  we  follow  the   procedure  described  in  Romero  et  al.
(2012). Briefly,  for each  sequence with a  given stellar mass  and a
thick  H$-$envelope,  as predicted  by  the  full  computation of  the
pre-white  dwarf evolution  (column 2  in Table  \ref{tabla-grid}), we
replaced   $^1$H  with   $^4$He  at   the  bottom   of   the  hydrogen
envelope. This is done  at high effective temperatures ($\lesssim 90\,
000$ K), so the transitory  effects caused by the artificial procedure
are  completely  washed  out  when  the  model  reaches  the  ZZ  Ceti
instability  strip.   The resulting  values  of  hydrogen content  for
different  envelopes are  shown in  Figure  \ref{envolturas-todas} for
each mass.  The  orange thick line connects the  values of $M_{\rm H}$
predicted   by   standard stellar   evolution   (column   2   table
\ref{tabla-grid}).  Note  that  the  maximum  value  of  the  hydrogen
envelope shows a strong dependence on the stellar mass. It ranges from
$2.4 \times 10^{-4} M_*$ for $M_* = 0.525 M_{\odot}$ to $1.4
\times 10^{-6} M_*$ for $M_* = 1.050 M_{\odot}$, with a value of $\sim
1  \times 10^{-4} M_*$  for $M_*  \sim 0.60  M_{\odot}$.  Computations
from  Romero et  al.   (2012)  are depicted  with  open circles  while
computations done  in this work  are indicated as filled  circles.  In
addition,  for   sequences  with   masses  between  0.878   and  1.050
$M_{\odot}$  and all  the values  of  the hydrogen  envelope mass,  we
computed three evolutionary sequences with different treatments of the
crystallization process  and the additional energy  sources related to
it:  (1) considering the  release of  latent heat  and the  release of
gravitational energy  due to phase separation using  the phase diagram
given by Horowitz  et al. (2010), (2) the same as  the former case but
considering the  phase diagram given by Segretain  \& Chabrier (1993),
and (3) considering the release of latent heat but neglecting chemical
redistribution  due to phase  separation.  Thus,  we can  consider the
crystallization treatment  as a kind  of extra parameter on  our model
grid.   To   our  knowledge,   this  is  the   first  time   that  the
crystallization process  is taken into account  self consistently with
these different prescriptions in  an asteroseismological study of white
dwarfs. Our goal is to use asteroseismology of massive white dwarfs to
ascertain   which   treatment   is   favored.    We   devote   Section
\ref{crystallization} to  study the effects of  crystallization on the
pulsation spectrum of massive  carbon$-$oxygen mass white dwarf stars,
and discuss the results  from our asteroseismological fits.  In closing,
we mention that by adding  the evolutionary sequences computed in this
work ($\sim 80$)  to those computed in Romero et  al.  (2012), we have
available a grid of $\sim 170$ evolutionary sequences, widely covering
the mass range  in which carbon$-$oxygen white dwarfs  are supposed to
be.

\subsection{Pulsation computations}
\label{pulsation-computations}

 We  computed nonradial  $g-$mode pulsations  of our  complete  set of
 massive  carbon$-$oxygen  core   white  dwarf  models employing  the
 adiabatic version of the LP-PUL  pulsation  code   described  in   C\'orsico   \&  Althaus
 (2006). The  pulsation code is based on  the general Newton$-$Raphson
 technique  that solves  the full  fourth order  set of  equations and
 boundary  conditions governing  linear, adiabatic,  nonradial stellar
 pulsations  following the  dimensionless  formulation of  Dziembowski
 (1971). We used the so  called ``Ledoux modified'' treatment to asses
 the run of the  Brunt$-$V\"ais\"al\"a frequency ($N$) (see Tassoul et
 al.  1990),  generalized  to  include  the effects  of  having  three
 different  chemical components  varying  in abundance.  This code  is
 coupled with the  LPCODE evolutionary code.  In order  to account for
 the effects of crystallization on the pulsation spectrum of $g-$modes
 we  have appropriately  modified the  pulsation code  considering the
 inner  boundary  conditions.  In  particular,  we  adopted  a  ``hard
 sphere'' boundary  condition, that assumes that the  amplitude of the
 eigenfunctions  of  $g-$modes  is  drastically  reduced  below  the
 solid/liquid interface  due to the non$-$shear modulus  of the solid,
 as compared  with the  amplitude in the  fluid region  (Montgomery \&
 Winget  1999). In  our code,  the  inner boundary  condition for  the
 population  of crystallized  models  is not  the  stellar center  but
 instead the  mesh$-$point corresponding to  the crystallization front
 moving  towards the  surface (see  C\'orsico  et al.  2004; 2005  for
 details).

The asymptotic period spacing is computed as in Tassoul et al. (1990):
\begin{equation}
\Delta\Pi_{\ell} = \frac{2\pi^2}{\sqrt{\ell (\ell +1)}}\left(\int_{r_1}^{r_2} \frac{N}{r}dr\right)^{-1}
\label{deltapi}
\end{equation}

\noindent where $N$ is  the Brunt$-$V\"ais\"al\"a frequency, and $r_1$
and  $r_2$ are  the  radii of  the  inner and  outer  boundary of  the
propagation  region, respectively. Note  that when  a fraction  of the
core  is  crystallized,  $r_1$   coincides  with  the  radius  of  the
crystallization  front, which  is moving  outwards as  the  star cools
down,  and the fraction  of crystallized  mass increases.   Hence, the
integral in  eq. (\ref{deltapi}) decreases, leading to  an increase in
the asymptotic period spacing, and also in the periods themselves.

\begin{figure*}
\begin{center}
\includegraphics[clip,width=350pt]{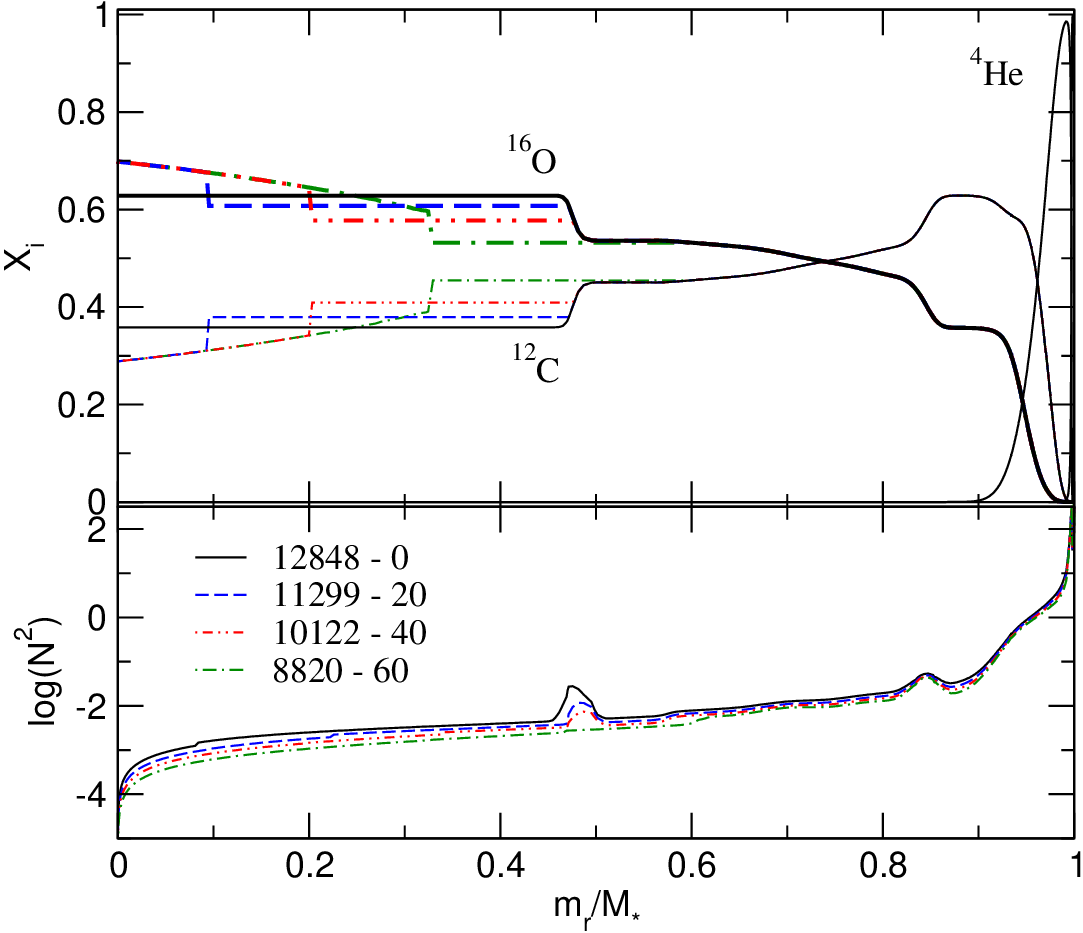}
\caption{Internal chemical profiles (upper panel)
and the logarithm of the square of the Brunt$-$V\"ais\"al\"a frequency (lower panel), 
in terms of the mass coordinate, corresponding to a DA white dwarf model
with $0.998 M_{\odot}$ at different stages of the cooling evolution. In this case, 
we consider the energy release of latent heat and phase separation upon 
crystallization according to Horowitz et al. (2010). Thick lines depict 
the oxygen chemical distribution, while thin lines depict the carbon abundance 
profile. For each model, the effective temperature and percentage of 
crystallized mass are indicated.}
\label{perfil-logq}
\end{center}
\end{figure*}

In Figure \ref{perfil-logq} we plot the inner chemical profiles (upper
panel)  and the  logarithm of  the square  of  the Brunt-V\"ais\"al\"a
frequency (lower panel),  for models with $M_* =  0.998 M_{\odot}$ and
different values  of effective  temperature. This sequence  belongs to
computations where  we employ the  phase diagram given in  Horowitz et
al.  (2010). The percentage of  crystallized mass present in the model
is indicated along with the effective temperature.

Each chemical transition region leaves  an imprint on the shape of the
Brunt-V\"ais\"al\"a  frequency, and  consequently  on the  theoretical
period  spectrum (Althaus  et al.   2003).   In the  core region,  the
presence of a  pronounced step at $m_r  / M_* \sim 0.49 $,  which is a
result  of the  occurrence of  extra$-$mixing episodes  during central
helium burning,  leads to a  narrow bump on  the Brunt$-$V\"ais\"al\"a
frequency profile.   When crystallization finally sets  in ($\sim 12\,
800$  K  for  this  stellar  mass),  rehomogenization  due  to  phase
separation modifies the structure  of the carbon and oxygen abundances
in the  central region.  As the crystallization  front moves outwards,
the  carbon$-$oxygen chemical  transition  at $m_r  /  M_* \sim  0.49$
becomes    smoother,    leaving    a    weaker    feature    on    the
Brunt$-$V\"ais\"al\"a   frequency.   The   feature  related   to  this
transition disappears for effective temperatures around $\sim 9\, 000$
K (see lower panel  of Figure \ref{perfil-logq}).  Additional features
in    the   Brunt-V\"ais\"al\"a   frequency    are   related    to   a
oxygen/carbon/helium  chemical transition  region, product  of nuclear
burning during the AGB and  thermally pulsating AGB stages, and to the
outer  helium/hydrogen  transition (see,  for  instance,  Figure 3  in
Romero  et  al. 2012).   These  chemical  interfaces  are modified  by
diffusion process acting during the cooling evolution.  In particular,
the oxygen/carbon/helium chemical transition is usually quite wide and
mode$-$trapping effects due to this  transition are not expected to be
too  strong, as  compared  to other  chemical  interfaces. The  outer
helium/hydrogen chemical transition is  also a source of mode trapping
associated with modes trapped in the outer envelope. The features induced on the Brunt-V\"ais\"al\"a by the oxygen/carbon/helium and helium/hydrogen chemical transitions will have a strong influence on the properties of the period spectrum. In particular, they determine the mode$-$trapping properties of DA white dwarf stars models  (Bradley \& Winget 1991, Brassard et al. 1992, C\'orsico et al. 2002).
Summarizing, we computed the  theoretical pulsation spectrum for about
$\sim  32\,600$ DA  white dwarf  models.  We  varied  three structural
parameters: the stellar  mass in the range $0.721  \leq M_* /M_{\odot}
\leq 1.050$, the  effective temperature in the range  $\sim 15\, 000 - 9\,
000$ K, and the thickness of  the hydrogen envelope in the range $-9.4
\leq \log (M_{\rm H}/M_*) \leq -4.5$, where the range of the values of
$M_{\rm H}$ depends  on the stellar mass.  In  addition, for sequences
with stellar  mass $\geq  0.878 M_{\odot}$, for  which crystallization
might occur at the effective temperatures considered, we computed the
theoretical pulsation spectrum  considering three different treatments
of  the  crystallization  process.  For  each model  we  computed  the
adiabatic oscillation  spectrum with harmonic  degrees $\ell =1$  and 2
and periods in the range 80$-$2000 s.

\section{Crystallization process in white dwarf stars}
\label{crystallization}
 
One of the additions we have done to the first model grid presented in
Romero  et  al. (2012)  is  to  include  different treatments  of  the
crystallization  process  in   our  computations.   Specifically,  for
sequences  with stellar  masses higher  than $\sim  0.878  M_{\odot}$, we
computed  the   white  dwarf  evolution   employing  three  different
treatments of  crystallization. i.e. by using the  phase diagrams from
Horowitz et  al. (2010) and Segretain  \& Chabrier (1993),  and by only
taking into account the release of latent heat as an additional energy
source. In this section we study the impact of crystallization process
on the theoretical pulsation spectrum  of massive DAV white dwarfs. We
analyze  the main differences  between the  considered crystallization
treatments   on  the   pulsation   properties  and   on  the   periods
themselves.

\subsection{Phase diagrams for dense carbon$-$oxygen mixtures}

As it is  known since Abrikosov (1960), Kirzhnitz  (1960) and Salpeter
(1961) a  white dwarf star  evolving on the cooling  track, eventually
will crystallize  as a  result of the  strong Coulomb  interactions in
their  very   dense  interior.  Crystallization  gives   rise  to  two
additional energy sources: latent heat (van Horn 1968) and the release
of gravitational energy due to phase separation in the carbon$-$oxygen
core (Garc\'{i}a$-$Berro et al.  1988ab).  As the oxygen$-$enriched solid
core grows at  the center of the white  dwarf, the lighter carbon$-$rich
liquid   mantle   left   behind   is  efficiently   redistributed   by
Rayleigh$-$Taylor  instabilities  (Isern et  al.   1997).  This  process
release gravitational energy, and  this additional energy source has a
substantial impact in the computed  cooling times of cool white dwarfs
(Segretain et al.  1994; Salaris et al.  1997; Montgomery et al. 1999;
Salaris et al.   2000; Isern et al. 2000; Renedo  et al. 2010; Althaus
et al. 2012).

For several years, the  standard phase diagram for crystallization used
in the stellar evolutionary computations of white dwarf stars was that
presented  in  Segretain  \&  Chabrier  (1993). These  authors  used  a
density$-$functional approach to obtain a phase diagram for an arbitrary
dense  binary  ionic mixture.   For  a  carbon$-$oxygen mixture  these
authors obtained  a phase diagram  of the spindle  type\footnote{For a
  spindle type phase diagram, the melting temperature of the mixture is
  always higher than that for  pure carbon, while in an azeotopic type
  phase diagram  the melting temperature  of the mixture can  be lower
  than that of pure carbon.},  strongly dependent on the charge ratio.
Recently,  the   phase  diagram  of   dense  carbon$-$oxygen  mixtures
appropriate  for white dwarf  star interiors  has been  re-examined by
Horowitz et  al. (2010).   This work was  motivated by the  results of
Winget et  al. (2009) who  found that the  crystallization temperature
for white dwarfs stars in  the globular cluster NGC 6397 was compatible
with the theoretical  luminosity function for  $0.5-0.535 M_{\odot}$ white
dwarfs and pure  carbon cores.  Horowitz et al.  (2010) used molecular
dynamic   simulations   involving   the   liquid  and   solid   phases
simultaneously,  allowing  a   direct  determination  of  the  melting
temperature and the composition of  the liquid and solid phases from a
single  simulation.  These  authors  predict an  azeotopic type  phase
diagram,  and   melting  temperature  considerably   lower  than  that
predicted by Segretain \& Chabrier (1993).  In fact, they conclude that
constraining the melting temperature of white dwarfs cores to be close
to that for pure carbon from Segretain \& Chabrier (1993) computations,
the oxygen  concentration should be around  $\lesssim 60\%$. Schneider
et al.  (2012) use the same technique as in Horowitz et al. (2010) based
on larger simulations with a larger number of ions and also included a
more accurate  identification of  liquid, solid and  interface regions
using a bond angle metric formalism.   As a result they obtain a phase
diagram close to  that obtained by Horowitz et  al. (2010).  Also, they
found an  excellent agreement  with the results  of  Medin  \& Cumming
(2010), who  used a semi$-$analytic  method to derive a  phase diagram
for multi$-$component plasma.   In particular, the differences between
the  results from  Horowitz  et al.  (2010)  and Medin  \& Cumming  et
al.  (2010)  for oxygen  abundances  near  $X_{\rm O}\sim  0.61-0.63$,
corresponding to the  $X_{\rm O}$ values for the  sequences with higher
stellar mass  ($\gtrsim 0.878 M_{\odot}$)  computed in this  work, are
small and we do not expect strong effects in our evolutionary
computation.   Finally,  Hugoto et  al.  (2012)  extend the  molecular
dynamic simulation technique  to a three component carbon$-$oxygen$-$neon
mixture  to  determine the  influence  of  $^{22}$Ne  on liquid  phase
equilibria. They found that the presence of a third component does not
appear  to impact  the chemical  separation found  previously  for two
component systems.

Althaus et al. (2012) presented  a detailed exploration of the effects
of  the new  phase diagram  given in Horowitz  et al.   (2010) on  the
evolutionary  properties of  white dwarfs,  and mainly  on  the cooling
ages.  They  employed the LPCODE  evolutionary code used in  this work
and initial accurate white dwarf structures derive from full evolution
of the progenitor  star from Renedo et al.   (2010), with stellar masses of 
$0.539$ and $0.878  M_{\odot}$.  These authors found that  for a given
stellar mass,  the amount of matter redistributed  by phase separation
is  smaller  when  the  Horowitz  et  al.   (2010)  phase  diagram  is
considered instead of the Segretain \& Chabrier (1993) one, leading to a
smaller  energy  release  from  carbon$-$oxygen  differentiation.   In
addition, the  composition changes are  less sensitive to  the initial
chemical profile. This  means that the magnitude of  the cooling delay
will  be  less affected  by  the  uncertainties  in the  carbon$-$oxygen
initial compositions and thus  by the uncertainties in the 
$^{12}$C$(\alpha , \gamma )^{16}$O reaction rate.

\subsection{The impact of crystallization on the pulsation spectrum}

For  this work  we  computed  the pulsation  spectrum  for dipole  and
quadrupole  $g-$modes for  model sequences with
stellar mass larger than  $0.878 M_{\odot}$ by employing both
Horowitz  et  al.  (2010)  (H2010)  and  Segretain  \& Chabrier  (1993)
(SC1993) phase  diagrams (see  Section \ref{model-grid}). In this way,  we can
study  the  impact  of  crystallization  on  the  adiabatic  pulsation
spectrum in general, and  the effects of the different crystallization
treatments on the pulsation properties, in particular.

As it is  well known, the temperature at  the onset of crystallization
depends  mainly  on  the stellar  mass,  as  can  be seen  from  Table
\ref{teff-crist}, where  we list  the atmospheric parameters  at which
crystallization begins,  for a given  stellar mass.  We only  show our
results for sequences with canonical hydrogen envelopes, meaning those
with  a $M_{\rm  H}$  value obtained  from  full stellar  evolutionary
computations.   As we  can  see, crystallization  begins  at a  higher
effective temperature for massive  sequences. This comes from the fact
that more massive white dwarf stars have higher central densities, and
since  the  crystallization  temperature  is proportional  to  density
($\sim \rho^{1/3}$  for a carbon pure composition),  it increases with
the  stellar  mass.   It  is  worth noting  that  the  crystallization
temperatures  when only  the  release  of latent  heat  is taken  into
account, are similar  to those for SC1993 phase  diagram.  In addition
to the dominant stellar mass dependence, there is a weak dependence of
the  crystallization  effective  temperature  with  the  mass  of  the
hydrogen envelope:  thinner hydrogen envelope  sequences usually begin
to crystallize at slightly higher effective temperatures. For example,
for  a $\sim  0.998  M_{\odot}$ model,  the crystallization  effective
temperature goes  from $12\,  773$ K  ($13\, 802$ K)  to $12\,  930$ K
($13\,  920$ K)  when $M_{\rm  H}$ decreases  from $M_{\rm  H}  = 1.98
\times  10^{-6}  M_*$ to  $M_{\rm  H}  =  4.96 \times  10^{-10}  M_*$,
considering   the   H2010   (SC1993)   phase   diagram   (see   Figure
\ref{crist-spec}).   Also,   for  a  given   stellar  mass,  different
crystallization  phase  diagrams  leads to  different  crystallization
temperatures, in agreement with the  results of Horowitz et al. (2010)
and  Althaus  et  al.  (2012)  (see  their Figure  1).  In  fact,  the
crystallization temperature  predicted using the  SC1993 phase diagram
is  $\sim  1000$ K  higher  than that  predicted  by  the H2010  phase
diagram.  Finally, note  that for sequences with a  stellar mass lower
than $0.878  M_{\odot}$ for SC1993,  and lower than  $0.917 M_{\odot}$
for H2010, crystallization begins at effective temperatures lower than
the observational red edge of the instability strip.

\begin{figure*}
\begin{center}
\includegraphics[clip,width=500pt]{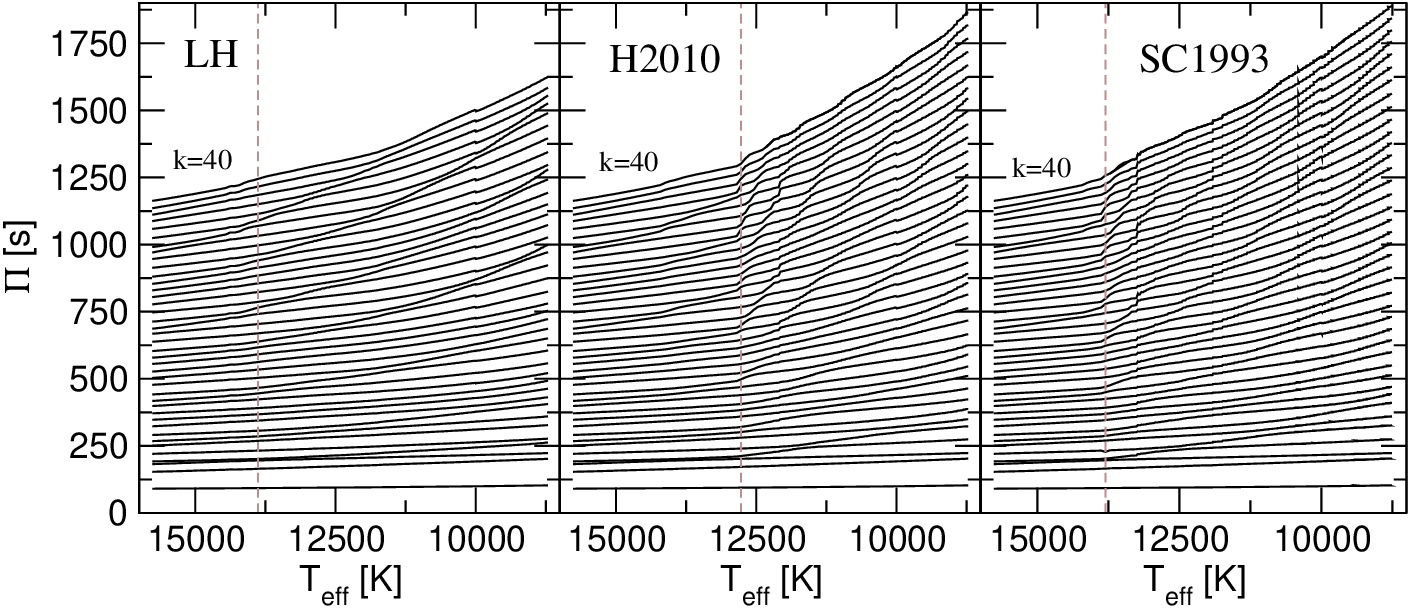}
\caption{Evolution of the periods for $\ell=1$ modes with $k=1-40$ as a function of the 
effective temperature, corresponding to a sequence with $M_* = 0.998M_{\odot}$ 
and canonical H envelope. Left panel: model where phase separation upon crystallization was neglected. 
Middle panel: result for H2010 phase diagram. Right panel: computation 
considering SC1993 phase diagram. The vertical line shows the value of $T_{\rm eff}$ 
when crystallization sets in.}
\label{crist-teff}
\end{center}
\end{figure*}

From  the results  presented in  Althaus et  al. (2012),  we  not only
expect differences  in the amount of  energy release, but  also on the
oxygen distribution in  the white dwarf interior (see  their Figures 2
and 3),  which in turn  should leave non$-$negligible imprints  in the
pulsation   spectrum.    We  start   by   analyzing   the  impact   of
crystallization  on  the   theoretical  period  spectrum.   In  Figure
\ref{crist-teff} we depict  the evolution of the $\ell  =1$ periods in
terms of  $T_{\rm eff}$ corresponding  to sequences with  stellar mass
$0.998M_{\odot}$ and canonical H envelopes, for the cases when we only
consider the  release of  latent heat as  an additional  energy source
(left panel, LH), and when we employ the H2010 (left panel) and SC1993
(right panel) crystallization  treatments. The vertical lines indicate
the effective  temperature at which crystallization  begins (see Table
\ref{teff-crist}). Note that low radial  order modes, $k=1, 2, 3$, are
almost   unaffected   when   we  include   phase  separation   upon
crystallization  in our computations.   On the  other hand  the period
values for modes  with radial order $k \geq 4$  begin to increase when
crystallization sets in the models.  This is because high radial order
modes behave  according to the  predictions of the  asymptotic theory,
where  the  period  values  are  given  by  $\Pi_k  \propto  k  \times
(\int_{r_1}^{r_2}|N|dr/r)^{-1}$  (Tassoul  et  al.  1990),  while  low
radial order modes do not follow the asymptotic prescriptions and then
are  not affected by  the crystallization  process.  In  addition, the
size of the propagation  region becomes smaller as the crystallization
front moves outwards  leading to an increase in  the period values. An
important feature  displayed in  Figure \ref{crist-teff} is  that when
crystallization  begins the  mode  bumping/avoided crossing  phenomena
propagates  to longer  periods, and  is  reinforced in  the H2010  and
SC1993 cases.
Since the crystallization temperature  is higher for SC1993 treatment,
at  a given  effective  temperature the  amount  of crystallized  mass
should  be larger  when  we employ  this  phase diagram  than when  we
consider the H2010  phase diagram.  In fact, for the $k  = 4$ mode and
$T_{\rm eff} = 11\, 600$ K,  the theoretical period value is 221.495 s
when we  only consider the release  of latent heat, and  238.164 s and
244.370 s when we also  take into account phase separation considering
the H2010 and the SC1993 phase diagrams, respectively.

\begin{figure*}
\begin{center}
\includegraphics[clip,width=350pt]{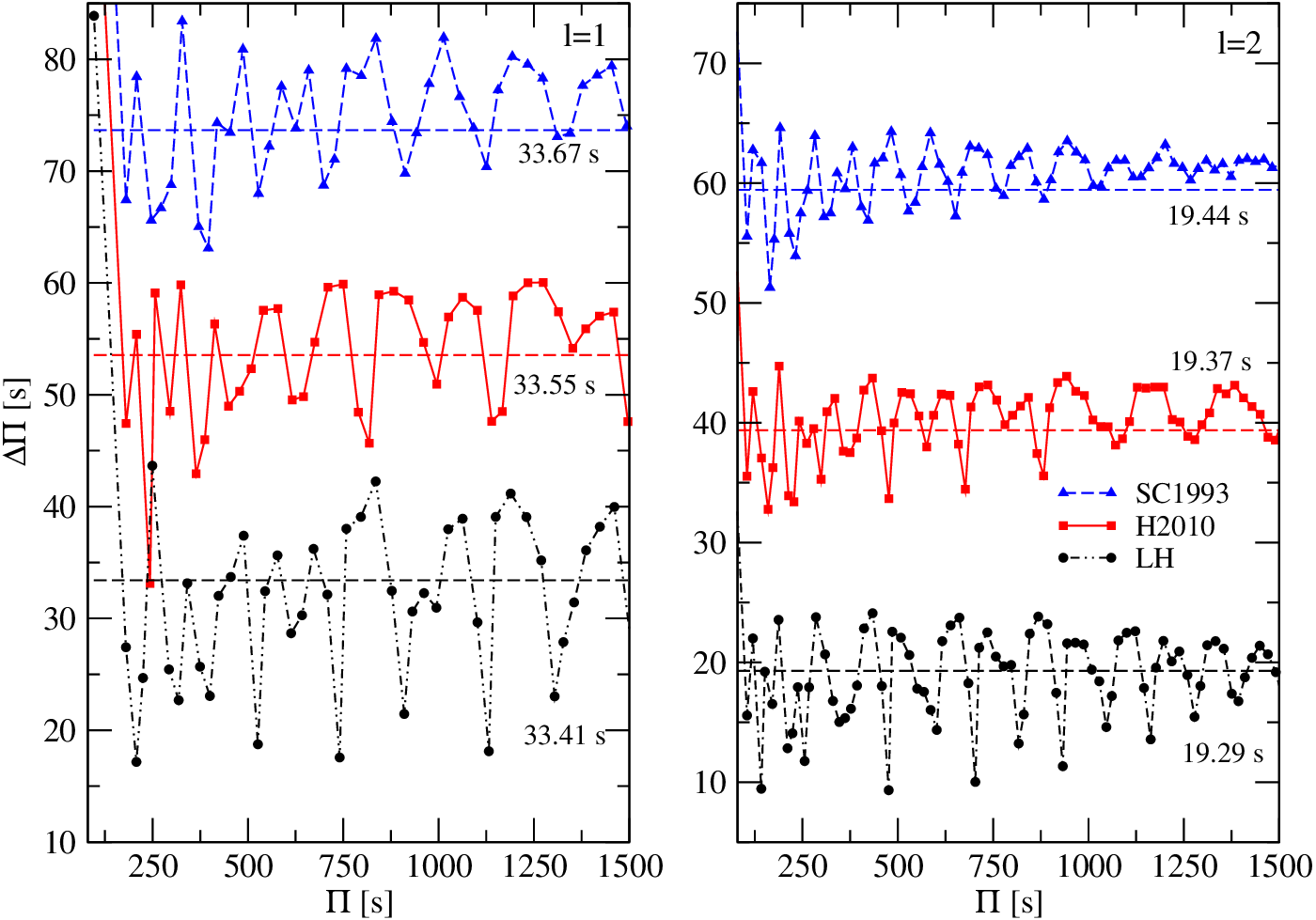}
\caption{Theoretical period spacing in term of the periods for models with 
$M_* = 0.998 M_{\odot}$ and $T_{\rm eff} = 11\, 600$ K for modes with 
$\ell =1$ (left panel) and  $\ell =2$ (right panel).  We show our results obtained by considering only the release of latent heat (black circles) and by employing the SC1993 (blue triangles) and H2010 (red squares) phase diagrams. The value of the asymptotic period spacing (horizontal lines) is indicated in each case.  The curves are shifted upwards in 20 s from the one below it, except for the bottom curve. }
\label{crist-deltap}
\end{center}
\end{figure*}

Finally,  in  Figure \ref{crist-deltap}  we  plot  the forward  period
spacing in  terms of  the periods, for  three models  characterized by
$M_* = 0.998  M_{\odot}$ and $T_{\rm eff} = 11\,  600$ K and different
crystallization treatments.   The left  panel depicts our  results for
$\ell  =1$ modes,  while  the right  panel  does it  for the  $\ell=2$
modes. Black circles  depict the results in the  case in wich we neglect phase
separation upon crystallization and only include the release of latent
heat in  our computations (LH),  while red squares and  blue triangles
depict the  results when we employ  the phase diagrams  of Horowitz et
al. (2010) and Segretain \& Chabrier (1993), respectively.

As the crystallization  front moves towards the surface  of the model,
not only the propagation region shortens but also the chemical structure of the
crystallized region becomes invisible to the oscillation.  This in turns will
leave a signature  on the periods spacing, as can  be seen from Figure
\ref{crist-deltap}.  Note  that the structure  of $\Delta \Pi$  has less
features when phase separation  upon crystallization is included in the
computations. The  minima in $\Delta  \Pi$ are less pronounced  and the
departure  from  the asymptotic  period  spacing (straight  horizontal
line)  is smaller  for the  H2010 and  SC1993 models  than for  the LH
model.  This  feature is more  noticeable for $\ell=2$ modes.   We 
also find these trends by comparing  the run of the period spacing for
the H2010 and SC1993 models. Since the computations using the Horowitz
et al.  (2010) phase diagram give a lower crystallization temperature
(see Table  \ref{teff-crist}), the percentage of  crystallized mass is 
$\sim  10  \%$ lower  for  the  H2010 model  than  for  the  SC1993
model. Then the structure of the period spacing is smoother
and shows less features for the SC1993 model.

In  closing, note  that the  value of  the asymptotic  period spacing slightly
increases with the amount of  crystallized mass in the models,
as predicted  by Eq.  \ref{deltapi}.  For $\ell  =1$ modes  this value
goes from 33.41  s when we neglect the release of  energy due to phase
separation, to 33.55  s when we employ the H2010  phase diagram in our
computations, and to  33.67 s when we consider  the phase diagram from
SC1993 instead.  This  increase, although small, is solely  due to the
change in the  treatment of crystallization, since we  are keeping the
stellar mass and effective temperature fixed.

\section{Stars analyzed and the spectroscopic mass}
\label{spectroscopy}

\begin{figure*}
\begin{center}
\includegraphics[clip,width=450pt]{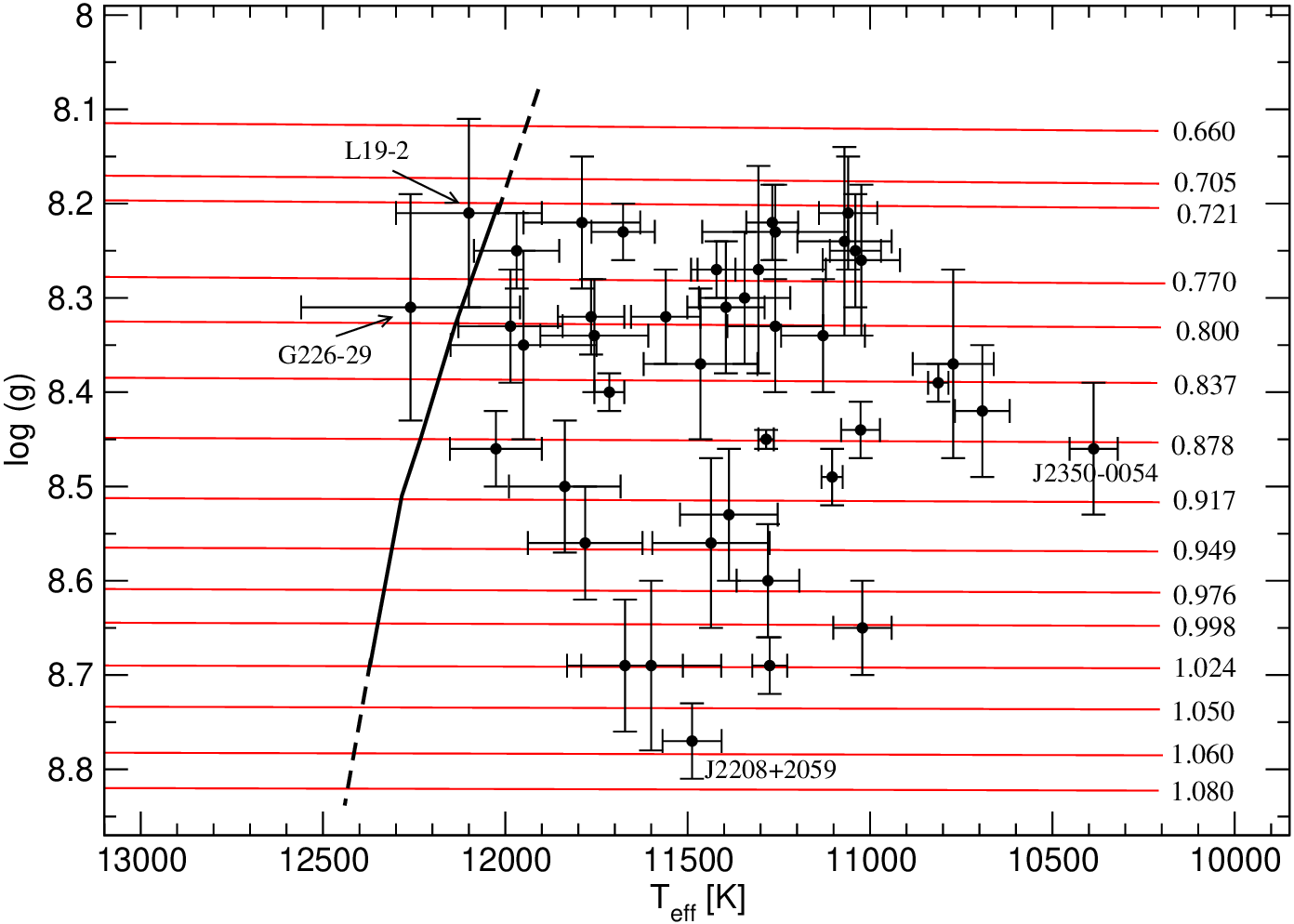}
\caption{The location of  the 42 ZZ Ceti stars  analyzed in this paper
  in the spectroscopic $\log g- T_{\rm eff}$ plane.  The thin red lines 
  correspond to our set
  of  DA white  dwarf  evolutionary tracks  with  thick (canonical)  H
  envelope thickness and stellar masses ranging from 0.660$M_{\odot}$ to 
  1.080$M_{\odot}$. The thick line depicts the theoretical blue edge of the 
instability strip for massive DA white dwarf stars  (see text for details).}
\label{gteff}
\end{center}
\end{figure*}

We analyzed a set of 42  ZZ Ceti stars with spectroscopic stellar mass
between 0.72$M_{\odot}$  and 1.05$M_{\odot}$.  This  sample belongs to
the massive  component of the white dwarf  mass distribution presented
in Kleinman  et al.  (2013). Note  that the mass  range considered for
the sample corresponds to white dwarfs thought to harbor cores made of
carbon and oxygen.   Thus, the most massive DAV  stars, like BPM 37093
($M_*  \sim  1.1 M_{\odot}$)  probably  having  an oxygen$-$neon  core
(Kanaan et  al. 1998; Metcalfe et  al. 2004; Kanaan et  al. 2005), are
not included  in our current  sample.  The atmospheric  parameters for
each  of  these  stars  are  listed  in  columns  2  and  3  of  Table
\ref{tabla-masas}.   For   the  first  36   objects,  the  atmospheric
parameters were  taken from Kleinman  et al. (2013), based  on spectra
taken  from the SDSS  Data Release  7 (Abazajian  et al.   2009).  For
J1916$+$3938 the  $T_{\rm eff}$  and $\log g$  values were  taken from
Hermes et al.  (2011). The last five objects are  bright ZZ Ceti stars
(e.g. Fontaine \& Brassard 2008),  four of which were already analyzed
from an  asteroseismological point  of view in  Romero et  al. (2012).
Since the classical ZZ Ceti stars have been targeted in several works,
there  are  several determinations  of  their atmospheric  parameters.
Thus,   within  the   uncertainties,  the   values  listed   in  Table
\ref{tabla-masas}  include all the  effective temperature  and surface
gravity determinations  from the literature.   
The fact that the uncertainties for the stars from the SDSS are smaller that those for the classical ZZ Ceti, may indicate that these uncertainties are most probably underestimated. However, we must note that SDSS counts with better flux calibrations and also that the $T_{\rm eff}$ and $\log g$ values are determined considering all the spectrum and not only the Balmer$-$line profiles as in Bergeron et al. (2004). The location of  the 42
DAV stars  targeted in this work on  the $\log g -  T_{\rm eff}$ plane
are shown  in Figure \ref{gteff},  along with the  evolutionary tracks
with masses ranging from 0.660 to 1.080 $M_{\odot}$.  Some objects are
indicated by their denomination. In particular, the two bright ZZ Ceti
stars G226-29 and L19-2, are the  hottest stars in the sample.  On the
other hand, J2350$-$0054  is the coolest massive DA  variable known to
date  (Mukadam  et  al.    2004)\footnote{The  DAVs  with  the  lowest
  effective  temperatures are  the low$-$mass  DAV stars,  supposed to
  have  helium core  (Hermes et  al.  2012,  Hermes et  al.  2013a).}.
Finally, J2208$+$2059 shows the  highest spectroscopic mass of our DAV
sample (Castanheira et al. 2013b).

Evolutionary sequences characterized by stellar masses in the range of
0.660 $-$  1.050$M_{\odot}$ correspond to  carbon$-$oxygen core models
(see  section \ref{model-grid}).  The sequences  with stellar  mass of
1.060 and  1.080$M_{\odot}$ have an oxygen$-$neon core  and were taken
from Althaus et al.  (2005c). The latter sequences were not considered
in our asteroseismological analysis.  It is important to note that all
the  sequences  were  generated  with  the  LPCODE  evolutionary  code
(Althaus et al.  2010b; Renedo et al.  2010).

The    spectroscopic    mass     values    (column    4    of    Table
\ref{tabla-masas}) were  estimated  by a  linear  interpolation of  the
evolutionary tracks  in the $\log g  - T_{\rm eff}$  diagram given the
values of $\log  g$ and $T_{\rm eff}$ inferred  from spectroscopy. The
mean value for the spectroscopic mass of our sample of 42 DAV stars is
$\langle  M_*  \rangle_{\rm  spec}   =  0.841  \pm  0.093  M_{\odot}$.
Kleinman et al.  (2013) found a  mean value of $ \langle M_* \rangle =
0.822 M_{\odot}$ for the massive component in the mass distribution of
DA  white   dwarf  stars  ($\sim   280$),  including   variable  and
non$-$variable  stars  but also  for  objects  with  masses above  1.050
$M_{\odot}$.  Taking  these differences into account,  and our limited
sample, the agreement between the  mean mass obtained in this work and
that of Kleinman et al. (2013) is excellent.

Finally, we include in Figure \ref{gteff} the theoretical blue edge of
the instability strip for massive  DA white dwarf stars, depicted as a
thick  vertical  line.   This  blue  edge was  obtained  by  means  of
non$-$adiabatic   computations  employing   the  non$-$adiabatic
  version  of  the  LP-PUL  pulsation  code  described  in  detail  in
  C\'orsico et al. (2006), adopting a MLT parameter $\alpha = 1.61$.
  Our computations  rely on the frozen  convection approximation, were
  the perturbation  of the convective  flux is neglected.   While this
  approximation is known to give unrealistic locations of the $g$-mode
  red edge  of instability, it  leads to satisfactory  predictions for
  the location of the blue edge of the ZZ Ceti (DAV) instability strip
  (see,  e.g., Brassard  \& Fontaine  1999),  for the  V777 Her  (DBV)
  instability  strip (see, for  instance, Beauchamp  et al.   1999 and
  C\'orsico  et al.   2009a), and  also for  the instability  strip of
  low$-$mass  pulsating  DAV  stars  (C\'orsico et  al.   2012c).   In
  addition, the stability  computations employing the time$-$dependent
  convection treatment  show that the  theoretical blue and  red edges
  are not dramatically  different from the ones found  by applying the
  frozen  convection approximation  (Van  Grootel et  al.  2012;  Saio
  2013). The location of the  theoretical blue edge of the instability
  strip is strongly dependent  of the convective efficiency adopted in
  the envelope  of the stellar models.  Then the location  of the blue
  edge can be hotter (cooler) if we adopt a larger (smaller) value for
  the mixing  length parameter  $\alpha$.  From Figure  \ref{gteff} we
  find that the location of  the theoretical blue edge agrees with the
  observations  since  most of  the  DAV  stars  are predicted  to  be
  pulsationally unstable. Although L19$-$2  and G226$-$29 are found to
  be hotter  than our theoretical blue edge,  within the uncertainties
  these objects are also inside the theoretical instability strip.

\section{New Observations}
\label{observations}

As part  of an  ongoing program devoted  to observe known  DA variable
white dwarfs, in order  to increase the  number of detected  modes, and
find new  variable ZZ Ceti stars,  we re-observed some  of the objects
from our sample  of massive DAV stars. Here we present new observations 
for five targets, obtained  at different campaigns from 2010 to 2013. 
All  targets were observed with
the Soar  Optical Imager and the Goodman  High Throughput Spectrograph
on the 4.1 m SOAR telescope,  in Chile. For details on the instruments
used  and the data  reduction see  e.g Castanheira  et al.  (2010). We
present a journal of observations in Table \ref{obs}.

\section{Results: Asteroseismological fits}
\label{fits}

For each  massive ZZ  Ceti star listed  in Table  \ref{tabla-masas}, we
search  for  an asteroseismological  representative  model, that  best
matches  the  observed  periods.   To   this  end,  we  seek  for  the
theoretical model that minimize the quality function given by Bradley
(1998):

\begin{equation}
\Phi (M_*, M_{\rm H}, T_{\rm eff})=\frac{1}{N}\sum_{i=1}^N  {\rm min}\left|\Pi_k^{\rm th} - \Pi_i^{\rm obs}\right|,
\label{phi}
\end{equation}

\noindent where  $N$ is the number  of periods observed  in the target
star.  Since the period spacing for $\ell =2$ is smaller than that for
$\ell =1$ modes,  there are always more quadrupolar  modes for a given
model when we consider a fixed period interval. So,  we require them to be  
closer to the observed  period by a
factor of $N_{\ell =2}/N_{\ell =1}$ in order to be chosen as a better
match (Metcalfe et al. 2004).

We also  considered the quality functions given  by C\'orsico et
al. (2009b):

\begin{equation}
\chi^2 (M_*, M_{\rm H}, T_{\rm eff}) = \frac{1}{N}\sum_{i=1}^{N} {\rm min}\left[ \Pi_k^{th} -  \Pi_k^{obs}\right]^2,
\label{chi}
\end{equation}

\noindent  and  the quality function employed in Castanheira \& Kepler (2008):

\begin{equation}
\Xi (M_*, M_{\rm H}, T_{\rm eff})=\sum_{i=1}^{N}\sqrt{\frac{\left[ \Pi_k^{th} -  \Pi_k^{obs}\right]^2 A_i }{\sum_{i=1}^N A_i}},
\label{xi}
\end{equation}

\noindent where the observed amplitudes  $A_i$ are used as weights for
each period. In  this way, the period fit is  more influenced by those
modes  with  large  observed  amplitudes.   Since  the  three  quality
functions  usually leads  to similar  results, we  shall  describe the
quality of our period fits in terms of the function $\Phi( M_*, M_{\rm
  H}, T_{\rm eff})$ only.

The results from our asteroseismological study are presented in Tables
\ref{tabla-periodos-1}    and    \ref{tabla-periodos-2}.   In    Table
\ref{tabla-periodos-1} we  list the results  for the 18  stars showing
three  or more  periods in  their observed  spectrum, while  for those
stars having one or two  observed periods the results are presented in
Table \ref{tabla-periodos-2}.   Both tables are  organized as follows.
In the  second, third and  fourth columns, we  show the values  of the
effective  temperature, stellar  mass  and thickness  of the  hydrogen
envelope for a  given asteroseismological model. Columns 5  and 6 show
the  observed  periods  and  amplitudes corresponding  to  each  star,
extracted from  different works listed in column  12.  The theoretical
periods,  along  with the  corresponding  harmonic  degree and  radial
order, are listed  in columns 7, 8 and 9,  respectively.  The value of
the quality function  $\Phi ( T_{\rm eff} , M_*,  M_{\rm H})$ for each
asteroseismological model  is listed  in column 10.   In column  11 we
list,  whenever  appropriate,  the  phase diagram  considered  in  the
treatment  of crystallization  and  the fraction  of the  crystallized
mass. For  several objects we  show more than  one asteroseismological
solution. The first  model listed is the one we choose  to be the best
fit  model for  that particular  object,  and refer  to the  remaining
solutions as the second and third solution, whenever it is the case.

In Table  \ref{tabla-sismology}  we  list  the  structural
parameters  of  the astroseismological  models  selected  as best  fit
models for  each star  analyzed in this  paper.  The  uncertainties for
$M_*$,  $T_{\rm  eff}$,  and  $\log  (L/L_{\odot})$  were  computed  by
employing  the following  expression  (Zhang, Robinson  \& Nather  1986;
Castanheira \& Kepler 2008)

\begin{equation}
\sigma^2_i = \frac{d_i^2}{(S-S_O)},
\end{equation}

\noindent where $S_0 \equiv \Phi  (M_*^0, M_{\rm H}^0, T_{\rm eff}^0 )$
is the minima of the quality function $\Phi$ reached at $(M_*^0, M_{\rm
  H}^0, T_{\rm eff}^0 )$ corresponding  to the best fit model, and
$S$ is  the value  of $\Phi$ when  we change  the parameter $i$  by an
amount $d_i$, keeping fixed  the other parameters.  The quantity $d_i$
can be evaluated as the minimum step in the grid of parameter $i$. The
uncertainties in the other quantities are derived from the uncertainties
in $M_*$, $T_{\rm eff}$, and $\log (L/L_{\odot})$. These uncertainties represent the internal errors of the fitting procedure. Other uncertainties come from the modeling itself. For example, the treatment of extra$-$mixing process depends on a free parameter $f$, that can vary at different stages of the evolution, and also will depend on the chemical composition of the convective region (see sec. \ref{input physics}). Also, the final carbon and oxygen central abundances depend on the value of the $^{12}C(\alpha, \gamma)O^{16}$ reaction rate, which cannot be estimated experimentally, leading to uncertainties on the final composition of the white dwarfs core. Finally, mass loss episodes depend on several parameters, including the metallicity, the amount of helium and rotation. Unfortunately a value for these uncertainties is not easy to asses.

Because most  of the stars in  our sample of  42 DAVs show only  a few
periods, we  can not  rely only  on the observed  periods to  select a
single asteroseismological  model, among all the  possible and equally
valid solutions, and we must apply some criteria. They are:

\begin{itemize}

\item First we looked for
those models associated  to minima in the quality  function, to ensure
that the  theoretical periods  are the closest  match to  the observed
values.  

\item When we  found several families of solutions  with similar values
  of  the quality  function, we  choose  those models  with values  of
  $T_{\rm eff}$ and $\log g$ as close as possible to the spectroscopic
  values.   In   particular,  we   consider   that  the   spectroscopic
  determination  of the  effective temperature  is more  accurate than
  that  of the surface  gravity, so we give more weight to the 
  spectroscopic value of $T_{\rm eff}$, as  we will  discuss in  the next
  section.

\item When possible, we  used the external identification of $\ell$
values for the observed periods,  mainly inferred from splitting due 
to the presence of rotation and/or magnetic fields, even if not all 
the components of the multiplet  reach observable amplitudes. 

\item When two or more modes have similar observed amplitudes in the 
power spectrum, we gave more weight to stellar models that fit those 
periods with theoretical periods having the same harmonic degree $\ell$.  

\item We give more weight  to solutions that fit the largest amplitude
  modes with  theoretical modes having $\ell =1$,  since dipolar modes
  would  exhibit  larger  amplitudes  than $\ell  =2$  modes,  because
  geometric cancellations effects become more important for modes with
  higher  harmonic  degree (Dziembowski  1977;  Robinson, Kepler  \&
  Nather 1982). However, for white dwarf stars having a large
  fraction  of  its  mass   in  a  crystallized  state,  the  possible
  propagation region for $g-$modes  is quite small, since oscillations
  cannot propagate  in the crystallized  regions of the star.  In this
  context, quadrupolar modes may be favored to be excited to observable
  amplitudes   since  they   have  shorter   wavelengths.   Thus,  this
  restriction can not be applied in these cases.

\end{itemize}

\subsection{Particular cases}

Next,  we   briefly   summarize    some   details   related   to   the
asteroseismological analysis for a few cases of interest.
\\

{\bf{J0048$+$1521}}: This star shows two modes with very close periods
at  615.3 and  604.19 s  with similar  amplitudes. In  this  case, the
period spacing will be of $\sim 11.11$ s only compatible with modes trapped
in  the outer  layers  (Althaus  et al.   2010b)  or with  unrealistic
massive white dwarf (Nityananda \&  Konar 2013; Das et al.  2013).  We
assume that  these two modes are  the $m=\pm 1$ components  of a $\ell
=1$ rotation triplet in which the central component is absent from the
pulsation  spectrum. For  our  asteroseismological fit  we consider  a
period at 609.75  s, corresponding to the $m=0$  component computed as
the average  value between the  two observed components.   By assuming
rigid slow rotation we infer a  mean rotation period of $\sim 9$ h, in
line  with   the  values  derived   for  other  ZZ  Ceti   stars  from
asteroseismology (see,  for instance, Table 4 of  Fontaine \& Brassard
2008).

{\bf{J0923$+$0120}}:  Variability  in  J0923$+$0120  was  reported  by
Mukadam  et al.  (2004),  with one  observed period  at $\sim  595$ s.
Further   observations   performed  during   2006   in  the   McDonald
Observatory, Apache Point, BOAO (Korea)  and HCT in India (A. Mukadam,
private communication)  reveled three additional very  long periods at
4145.0, 2032.3 and  1436.37 s. The longest period  was dismissed since
it is most  likely to be low frequency  noise.  Surprisingly, no modes
with   periods  between   600  and   1400  s   were  found.    In  our
asteroseismological study  we consider  only two periods,  595.055 and
1436.37  s.  Because  the theoretical  periods computed  in  this work
reaches a longest value of $\sim$  2000 s, the period at 2032.30 s was
not included in our analysis.

{\bf{J1323$+$0103}}:  This  star   shows  the  most  populated  period
spectrum   of  our  sample,   with  15   periods  considered   in  our
seismological fit.  The  periods listed in Table \ref{tabla-periodos-1},
are  a  combination  of  two  sets of  observations.   The  first  set
corresponds to  those periods reported  by Kepler et al.   (2012), who
performed asteroseismological fits  employing the model grid presented
in  Romero et  al.   (2012)  and Castanheira  \&  Kepler (2008).   The
results  from  seismology  by   using  the  full  evolutionary  models
following Romero et  al. (2012) are: $M_* =  0.88 \pm 0.02 M_{\odot}$,
$T_{\rm eff} = 12\, 100 \pm 140$  K, $M_{\rm H} = (4.0 \pm 3.3) \times
10^{-7} M_*$, $M_{\rm He} = (2.6 \pm 0.3) \times 10^{-3} M_*$, $X_{\rm
  C}  = 0.37  \pm 0.01$ and  $X_{\rm  O} =  0.62 \pm  0.01$. The  results
obtained by  employing the models  described in Castanheira  \& Kepler
(2008), which  assume a central composition  C/O = 50\%  and allow the
hydrogen and helium  layer mass to vary, are: $T_{\rm  eff} = 11\, 900
\pm 200$ K, $M_* = 0.88 \pm 0.08 M_{\odot}$, $M_{\rm H} = 10^{-4.5 \pm
  0.4} M_*$  and $M_{\rm  He} = 10^{-2.3  \pm 0.5} M_*$.   The large
difference in the  hydrogen content of the two  seismological fits can
be interpreted in terms of the core$-$envelope symmetry (Montgomery et
al.    2003)  and   the   differences  in   the  chemical   structures
characterizing both model grids.

A second  set of periods  was obtained from observations  performed at
the  SOuthern Astrophysical  Research  telescope (SOAR)  in 2012  (see
Sec. \ref{observations}).  Although some modes were already present in
the first set,  most of the observed modes in the  second set have not
been  observed before.  In our  asteroseismological study  we consider
both set of observed periods, using an average value for those periods
present in  both sets.   We obtain a  best fit model  characterized by
$M_* = 0.917\pm 0.020 M_{\odot}$, $T_{\rm eff} = 11\, 535 \pm 72$ K, $M_{\rm H}
= 3.90\times  10^{-6} M_*$,  $M_{\rm He} =  1.31 \times  10^{-3} M_*$ and
$X_{\rm  O} =  0.609$.  The  stellar mass  is a  bit higher  than that
obtained in the previous  seismological study, but compatible with the
spectroscopic  determinations.  On  the other  hand, the  mass  of the
hydrogen  layer is  about one  order  of magnitude  thicker than  that
obtained by employing the models from Romero et al. (2012),  but still 
thin as compared with the
solution obtained using the models of Castanheira \& Kepler (2008).

{\bf{J1612$+$0830}}:  This  star  was   reported  to  be  variable  by
Castanheira  et al.  (2013a),  with two  very close  observed periods,
115.17 and  117.21 s, and  very similar observed amplitudes.   A close
inspection of  the Fourier transform  (see Figure 1 of  Castanheira et
al. 2013a) shows  the presence of a low amplitude  third peak at $\sim
112$ s.  Therefore, the three observed periods are the components of a
rotational $\ell =1$ triplet, being the central component at 115 s. We
employ the  two main components at  115 s and  117 s to derive  a mean
rotation period of $\sim 1$  h, considering slow rigid rotation.  Note
that J1612$+$0830  shows a $\ell=1$  triplet with very  short periods,
similar to  G226$-$29. Also  the seismological solutions  obtained for
both  stars are  very  similar,  with a  thick  hydrogen envelope  and
effective temperatures close to  the blue edge.  However, the rotation
period derived for G226$-$29 is $\sim 9$ h (Kepler et al. 1995).

{\bf{J1711$+$6541}}: As  J0048$+$1521, this star shows  two modes with
very close  periods at 612.6 and  606.3 s and  similar amplitudes.  We
assume that  these modes are the  $m=\pm 1$ components of  a $\ell =1$
rotation triplet,  with a  not observed $m=0$  component at  609.45 s.
Considering rigid, slow  rotation we derive a mean  rotation period of
$\sim 16.4$ h.

{\bf{J2128$-$0007}}: Although the  spectroscopic mass for J2128$-$0007
is close to the lower limit  of our sample, $\sim 0.788 M_{\odot}$, we
obtain   a   seismological   solution   with   a   stellar   mass   of
0.976$M_{\odot}$.   The second  solution with  $\sim  0.593 M_{\odot}$
listed  in  Table  \ref{tabla-periodos-1}  is probably  related  to  the
presence of the two modes with  periods at $\sim$ 274 s and $\sim$ 304
s, since they are close to the modes observed in G117$-$B15A at 270.46
s and  304.05 s (Kepler et al.   1982). Romero et al.   (2012) found a
seismological solution for G117$-$B15A with the same stellar mass, and
a  thin  H  envelope  ($M_{\rm  H} =1.25  \times  10^{-6}  M_*$).   For
J2128$-$0007 we  obtain a solution with thick  hydrogen envelope, since
for G117$-$B15A this parameter is  basically set by the mode at $\sim$
215 s (see Romero et al. 2012 for details).

{\bf{J1916$+$3938}}: This  star is the first pulsating  DA white dwarf
star  located in  the  {\it  Kepler} mission  field  ({\it Kepler}  ID
4552982) and  identified through ground$-$base  time series photometry
by  Hermes et  al.  (2011). As  a  result, these  authors found  seven
possible modes that are listed in  Table 5. The first seismological study 
applied to this object was performed by C\'orsico et al. (2013b) employing 
the model grid presented in Romero et al. (2012). In this work we reanalyze 
this star and obtain the same asteroseismological model.
Note that the two shortest
modes,  with  periods  at  823.9  and  834.1  s  are  associated  with
theoretical   modes   showing   different  harmonic   degrees.   Other
possibility is that these two modes  are the $m = \pm 1$ components of
a rotational triplet in which the central component is not present. The
period of the missing $m=0$  component can be estimated as the average
of  the $m=\pm  1$ components,  at 829  s. Under  this  assumption, we
performed an asteroseismological fit replacing the two shortest periods
by their average. As a result we obtained the same asteroseismological
solution as before.   Finally, by assuming rigid and  slow rotation we
can infer a mean rotation period of 18.77 h.

{\bf{The  classical ZZ  Cetis:  G226$-$29, L19$-$9,  G207$-$9 and  BPM
    30551}}: of our sample of 42  massive DAV stars, four of them were
also part of the sample studied  by Romero et al. (2012). Since now we
have available an expanded grid, and the additional parameter given by
the crystallization treatment, we  think it is worthwhile to reanalyze
these objects.  The observed periods and amplitudes are the same as in
Romero et  al. (2012),  except for L19$-$2  that was  re-observed (see
Sec.   \ref{observations}).  The  asteroseismological  models obtained
for G226$-$29 and  G207$-$9 are the same as  those presented in Romero
et al. (2012).  For  BPM 30551 we obtain a best fit  model with $M_* =
0.721 M_{\odot}$ and  $T_{\rm eff} = 11\, 157$ K  and $\Phi =0.176$ s,
in  addition to a  second solution  with $M_*  = 0.705  M_{\odot}$ and
$T_{\rm eff}  = 11\, 436$ K  and $\Phi =0.175$ s,  that corresponds to
the  seismological solution  found in  Romero et  al. (2012)  for this
star.  We select the seismological  model with $M_* = 0.721 M_{\odot}$
over  the model  with $M_*  = 0.705  M_{\odot}$ because  the effective
temperature  and   stellar  mass  are  in  best   agreement  with  the
spectroscopic  values. Finally,  for  L19-2 we  found a  seismological
solution  corresponding to  the  same evolutionary  sequence than  the
solution presented in Romero et al. (2012) but $\sim 70$ K cooler.

\subsection{Very cool massive ZZ Ceti or low mass variable white dwarfs?}

From  our sample of  42 DAV  stars, we  focus here  on three  of them:
J0000$-$0046  reported  by  Castanheira  et al.  (2006),  J0940$+$0052
discovered by Castanheira et  al. (2013a) and J2350$+$0054 reported by
Mukadam et  al. (2004). From  spectroscopy, these stars show  very low
effective  temperatures and high  surface gravities,  being J2350+0054
the most extreme, with $T_{\rm eff} =10\,  387 \pm 66$ K and $\log g =
8.46 \pm 0.07$. Since these stars are located near the red edge of the
instability strip, they should have a period spectrum characterized by
several modes with  long periods (Mukadam et al.  2006).  However, 
the three  objects show only a  few modes with short  periods in their
observed spectrum, characteristic  of DAV stars near the  blue edge of
the instability strip.  This is particularly true for J0940$+$0052 and
J2350$+$0054 that show a few modes with periods between 250 and 400 s.
Here,  we   consider  two   possible  simple  explanations:   (1)  the
spectroscopic  determination  of  the  effective  temperature  is  not
correct and these stars are  hotter than predicted, or (2) the surface
gravity values are  not correct and these stars  are not massive white
dwarfs  but low  mass white  dwarf stars,  that are  also known  to be
variable (Hermes et al. 2012, 2013a). Since usually the $\log g$ value
is the  one that  is most poorly  determined from spectroscopy,  as we
discuss  in  Section \ref{stellar-mass-ast},  we  consider the  second
possibility as  the most likely.   For low mass variable  white dwarfs
the  blue edge  of the  instability strip  is located  at considerably
lower effective temperatures than for  the classical ZZ Ceti stars, as
shown  in C\'orsico  et al.   (2012c).  In  fact, if  we  consider the
values of $T_{\rm eff}$ for J0000$-$0046, J0940$+$0052 and J2350+0054,
we find that  they are compatible with stars located  close to the low
mass stars blue edge, with  stellar masses $\gtrsim 0.4 M_{\odot}$ for
the   first   two   objects,   and   $\gtrsim   0.3   M_{\odot}$   for
J2350$+$0054. Further analysis on these objects needs to be done, both
from the  observational as well as from  the asteroseismological point
of view, but with more modes detected.

\subsection{Stellar mass from asteroseismology}
\label{stellar-mass-ast}

\begin{figure}
\begin{center}
\includegraphics[clip,width=230pt]{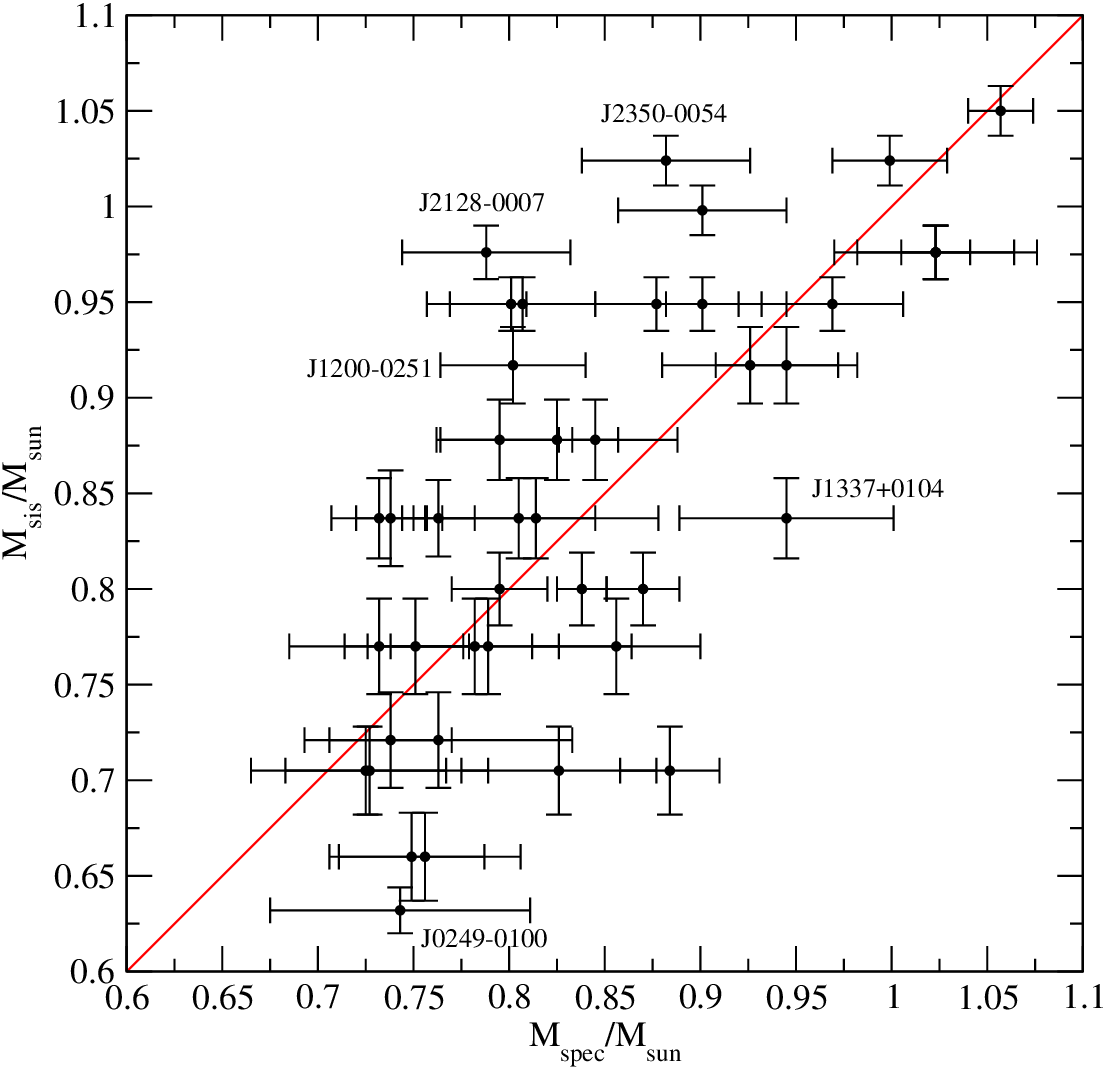}
\caption{Comparison between the values of the stellar mass according 
to spectroscopy and asteroseismology. The uncertainties in the 
asteroseismological mass are the internal uncertainties of the fitting 
procedure. The red line 
represents the 1:1 correspondence.}
\label{mass-correlation}
\end{center}
\end{figure}

In this section  we compare the results obtained  for the stellar mass
from  spectroscopy and  seismology for  our  sample of  42 DAV  stars.
Since  the  evolutionary models  employed  to  obtain a  seismological
representative model  for each object are  the same we use  to derive the
spectroscopic mass from the observed atmospheric parameters, this
comparison  is worth  doing.  In  Figure \ref{mass-correlation}  we  compare both
determinations of the  stellar mass.  The red diagonal  line shows the
1:1 correspondence.   As we  can see from  this figure,  the agreement
between both determination is not very good, reaching discrepancies as
high as $\sim 0.2  M_{\odot}$ for J2128$-$0007 and J1200$-$0251, $\sim
0.14 M_{\odot}$  for J2350$-$0054 and around $\sim  0.1 M_{\odot}$ for
J1337+0104.  On  the other  hand the bulk  of the points  does cluster
around the  1:1 correspondence  line, implicating that  no significant
offset  is present between  the asteroseismological  and spectroscopic
determinations  of the  stellar  mass.  

There  are some shortcomings that can  lead to erroneous
determination of the observational  parameters and thus affect further
determinations of the stellar mass  from spectroscopic data as well as
from  asteroseismology.  As  it is  well known,  determination  of the
surface gravity for  cool  DA white  dwarf  stars with  effective
temperatures $\lesssim 12\, 500$ K  leads to higher values of $\log g$
than those of hotter stars. First  it was thought that the presence of
helium from convective mixing could  mimic the effects of high surface
gravities (Bergeron et al.  1991). However, this possibility was ruled
out by  Tremblay et al. (2010)  who analyzed the high  S/N spectrum of
six  DA white dwarfs  with effective  temperatures between  $\sim 12\,
500$ K and $\sim 10\, 500$ K, and found no traces of helium.  The most
popular  alternative  explanation  is  related  to  the  treatment  of
convective  energy transport,  which is  currently represented  by the
mixing length  theory in 1D models  (Tremblay et al.  2010; Koester et
al.  2009).  In  addition, Tremblay et al.  (2011,  2013) show that 3D
model  spectra provide  a  much better  characterization  of the  mass
distribution of  white dwarfs  and the shortcomings  in the  1D mixing
length theory was in fact producing the high$-\log g$ problem.

Also, the presence of a  magnetic fields can mimic the line broadening
produced by a  high surface gravity in the  observed spectrum when the
resolution in wavelength  is not good enough to  resolve the different
components of the splitting, leading to an incorrect (higher) $\log g$
determination (Kepler et  al. 2013).  On the other  hand, the presence
of  a magnetic  field can  inhibit the  development of  certain modes,
and/or  its components, if  the displacement  direction of  the moving
material  is  perpendicular  to   the  magnetic  field  (Arras  2006).
Therefore  some normal  modes could  not  be present  in the  observed
pulsation spectrum of the star.

\begin{figure}
\begin{center}
\includegraphics[clip,width=220pt]{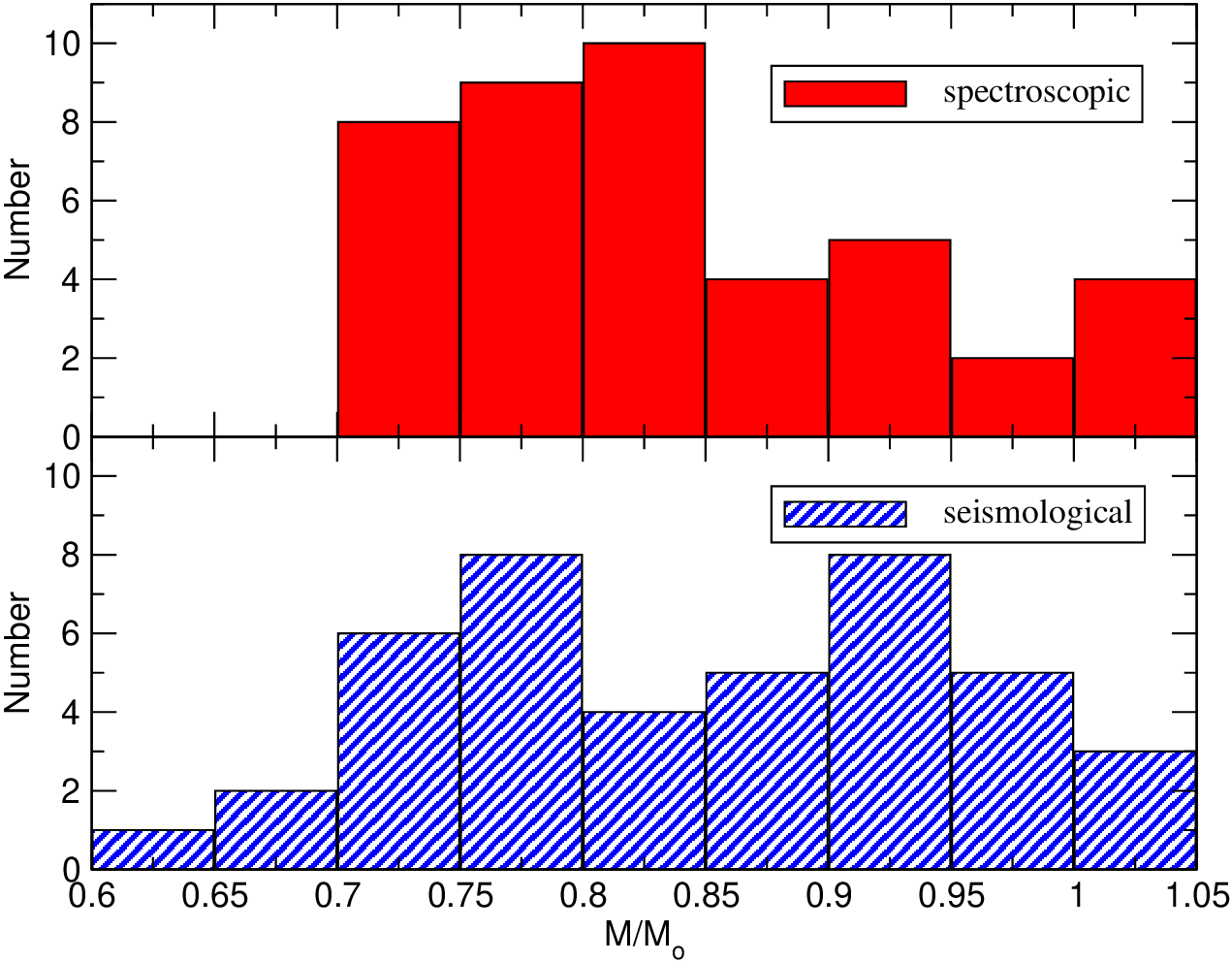}
\caption{Histograms depicting the mass distribution for the sample 
of 42 ZZ Ceti massive stars with carbon$-$oxygen core, studied in this
 work, according to our spectroscopic (upper panel) and seismological 
(lower panel) analysis. }
\label{histograma-mass}
\end{center}
\end{figure}

Figure \ref{histograma-mass}  shows the  mass distribution for  the 42
massive ZZ Ceti stars analyzed  in our work, according to spectroscopy
(upper  panel) and asteroseismology  (lower panel).  The spectroscopic
mass distribution has  its main contribution from the  low mass range,
for  stellar masses  below  $0.85M_{\odot}$. On  the  other hand,  the
sesimological mass distribution shows two components, between $ 0.75 -
0.8 M_{\odot}$  and $0.9  -0.95 M_{\odot}$. This  could be due  to the
specific values of stellar mass in  our model grid.  The mean value of
the  asteroseismological mass is  $\langle M_*  \rangle_ {\rm  seis} =
0.850  \pm 0.110  M_{\odot}$, slightly  larger than  the spectroscopic
value $\langle M_*  \rangle_ {\rm spec} = 0.841  \pm 0.093 M_{\odot}$.
Note that  the methods employed to  derive both values  of the stellar
mass are quite different. Also, they rely on two different independent
sets of  observational data, being the spectrum  for the spectroscopic
mass,  and the  observed  periods  in the  case  of the  seismological
determinations. Taking  this fact into account,  the agreement between
the spectroscopic and the seismological mean mass is satisfactory.

\subsection{Effective temperature from asteroseismology}
\label{teff-ast}

\begin{figure}
\begin{center}
\includegraphics[clip,width=230pt]{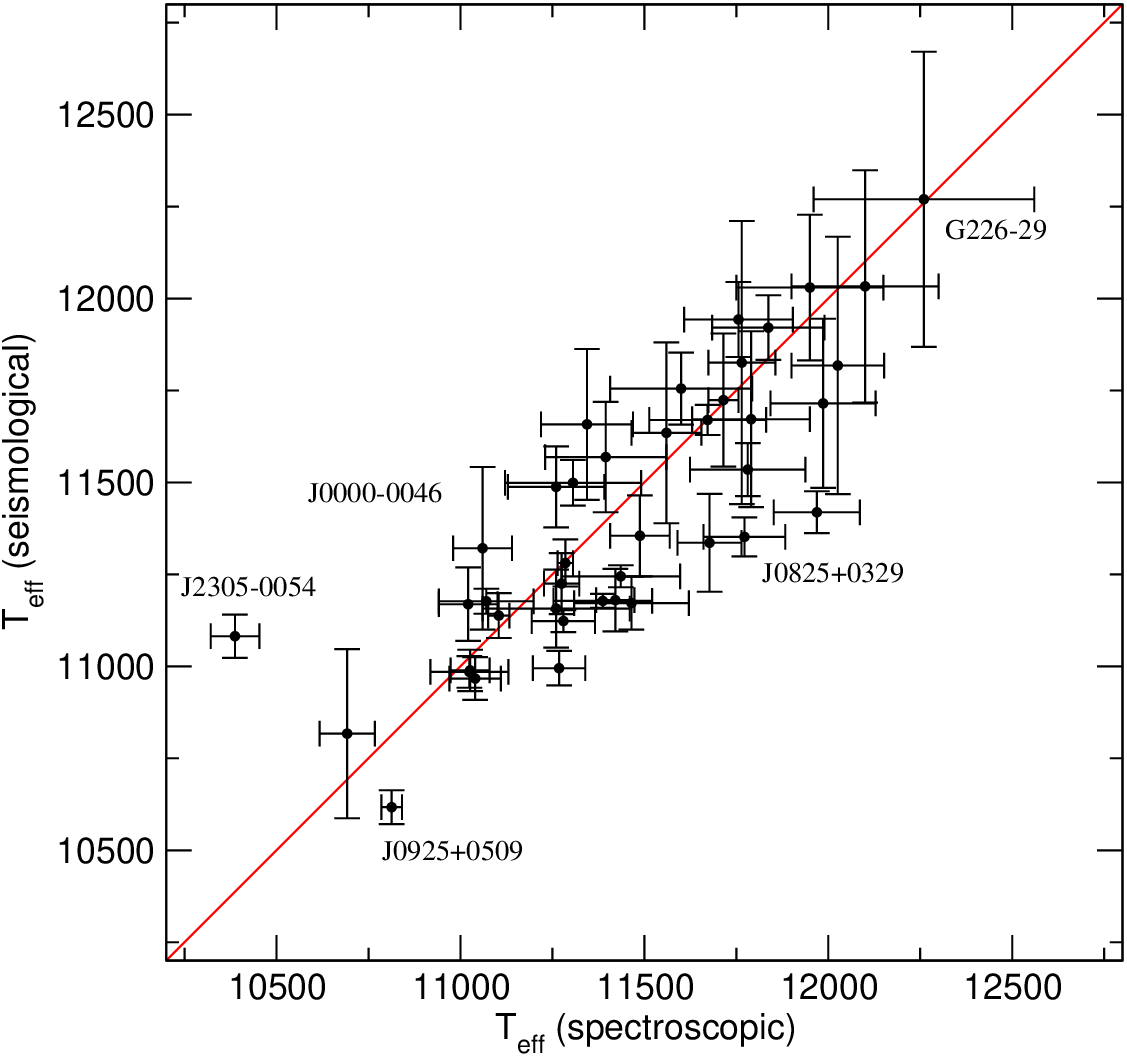}
\caption{Comparison between the spectroscopic ($x$ axis) and 
seismological ($y$ axis) determination of effective temperature 
for our sample of 42 DAV stars. The red line indicates the 1:1 
correspondence.}
\label{teff-correlation}
\end{center}
\end{figure}

In   Figure  \ref{teff-correlation}   we  compare   the  spectroscopic
($x-$axis) and  asteroseismological ($y-$axis) determinations  for the
effective  temperature. Error bars  in the  asteroseismological values
depict the  internal uncertainties  from the fitting  procedure, while
the   uncertainties  in  the   spectroscopic  determination   are  the
spectroscopic  fitting  uncertainties  (see Table  \ref{tabla-masas}).
The diagonal  red line shows the  1:1 correspondence.  As  can be seen
from  this figure  the correspondence  between both  determinations is
quite good,  specially in the high effective  temperature domain.  The
larger   discrepancies  appear   to  be   located  at   low  effective
temperatures.  In particular,  for J2305$-$0054, and J0000$-$0046, the
seismological effective temperature  is $\sim 700$ K and  $\sim 580$ K
higher   than   the    spectroscopic   value,   respectively.    Large
discrepancies at  low temperatures  are expected since  the atmosphere
models  employed to fit  the observed  spectra use  the MLT  theory of
convection.  The MLT theory is a good approximation for stars near the
blue edge of  the instability strip because the  outer convective zone
is still very thin.   However, for lower effective temperatures closer
to the red  edge of the instability strip  the outer convection region
is quite  thick, and the shortcomings  from applying the  MLT might be
important.  In   addition,  the  convective   properties  do  change
considerably in the effective temperature range $13\, 000 -6\, 000$ K,
as  is  seen  from  3D  model  atmosphere  computations  (Tremblay  et
al. 2013).

As we  mention, one of the  criteria applied to  elect a seismological
model  as a  possible solution  is set  by the  atmospheric parameters
determined from  spectroscopy. In particular,  we give more  weight to
the effective  temperature determination than the  $\log g$, therefore
our seismological solutions  tend to have a better  agreement with the
spectroscopic  effective  temperatures  than  with  the  spectroscopic
stellar masses, as can be seen from Figures \ref{mass-correlation} and
\ref{teff-correlation}. If  we gave  more weight to  the spectroscopic
surface  gravity,  the   differences  between  the  spectroscopic  and
seismological values of $T_{\rm eff}$  would be as large as $\sim 500$
K for $T_{\rm eff} \gtrsim 11\, 000$K.

\subsection{The thicknesses of the Hydrogen envelope}
\label{mh-ast}

\begin{figure}
\begin{center}
\includegraphics[clip,width=220pt]{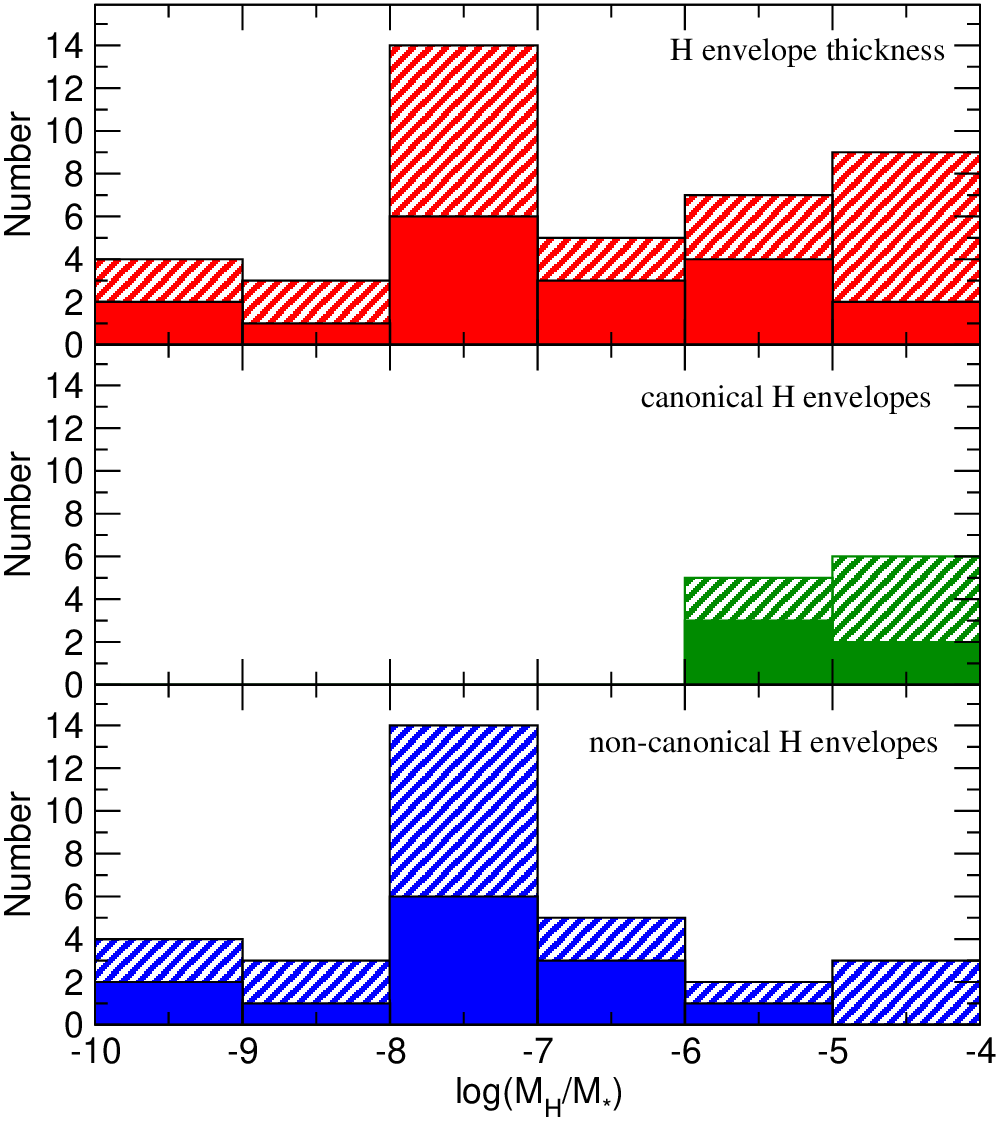}
\caption{Upper   panel:  histogram   showing  the   hydrogen  envelope
  thickness   distribution for   our complete  sample  of   42  massive   
DAV  stars (dashed).  Middle  panel: histogram for models with  canonical 
values of
  the hydrogen  envelope thickness,  as predicted by  stellar evolution
  theory  for  each  mass.  Lower  panel: histogram  for  models  with
  non$-$canonical values of the hydrogen envelope thickness, as obtained
  by  means   of  our   artificial  procedure  described   in  Section
  \ref{model-grid}. We only consider  the best fit model solutions for
  each star. In adition, with fill bars, we show the results corresponding 
to a sample composed only with stars having 
three or more modes in their spectrum (see Table \ref{tabla-periodos-1}). }
\label{env-histograma}
\end{center}
\end{figure}

Although there  is some observational  evidence of the existence  of a
range  in  the  thickness  of  the hydrogen  envelope,  currently  the
hydrogen content can be  determined only by asteroseismology. Since we
analyze a large  number of DAV stars, we can shed  some light over the
distribution of  the hydrogen  envelope mass.  In  the upper  panel of
Figure \ref{env-histograma}  we present our  results for the  42 stars
analyzed in this  work (dashed bars). We only show  the best fit model
for  each object.   The  distribution shows  two  maximums: for  thick
hydrogen envelopes, in the $\log  (M_{\rm H}/M_*)$ range $-$5 to $-$4,
and for thin values in the range $-$7 to $-$8.  In the middle panel of
Figure \ref{env-histograma} we show the histogram corresponding to the
asteroseismological models having  canonical envelopes, that amount to
10 stars.  Note  that the maximum amounts of  hydrogen as predicted by
canonical evolutionary computations, which depends on the stellar mass
value, are in the range $-$4  to $-$6 for stellar masses considered in
our grid.   Then, the peak for  thick envelopes is  mostly composed by
models with  canonical envelopes. Finally,  in the lower panel  of the
figure we present the histogram for the seismological models showing a
non$-$canonical  envelope thickness, that  is, envelopes  thinner than
those predicted  by our standard stellar evolution  models.  We recall
that  these  thinner  envelopes  where  generated  via  an  artificial
procedure described in Section \ref{model-grid} in order to extend the
exploration of the parameter space of the models for asteroseismology.
As it  is expected, the second  peak in the distribution  in the range
$-$7  to $-$8 is  completely composed  by models  with non$-$canonical
hydrogen envelopes,  that amount to  14 objects.  Also, it  appears to
exist a  much less notorious third  peak in the  distribution for very
thin envelopes in  the range $-$10 to $-$9.  
The  envelope  distribution for  a  reduced  sample,  composed by  the
objects   showing   three   or   more  observed   modes   (see   Table
\ref{tabla-periodos-1}),  is depicted  in  Figure \ref{env-histograma}
with  filled symbol. As  can be  seen from  this figure,  the envelope
distribution has a similar shape when compared with the distribution for
 the full sample. There is a dominant peak in the range $-7$ to $-$8
and a weaker  contribution from very thin envelopes  in the range $-9$
to  $-$10, both  composed by  non$-$canonical envelopes.  In addition,
five out  of six seismological  models with hydrogen envelopes  in the
range $-$4 to $-$6 correspond to canonical models. In this case, 72 \% 
of the seismological models have non$-$canonical envelopes.
Romero et al.  (2012), using a different  sample of
DAV stars  characterized with  stellar masses around  $0.6 M_{\odot}$,
also found a peak  in the hydrogen envelope distribution corresponding
to very thin envelopes (see  their Figure 11), apart from the dominant
component  in   the  range  $-$4  to  $-$5.    The  hydrogen  envelope
distribution presented in that work  does not show a thin component in
the range $-$7  to $-$8. Finally, note that  most of the seismological
models  obtained in  our  study, $\sim  76  \%$, have  non$-$canonical
envelopes.

\begin{figure}
\begin{center}
\includegraphics[clip,width=250pt]{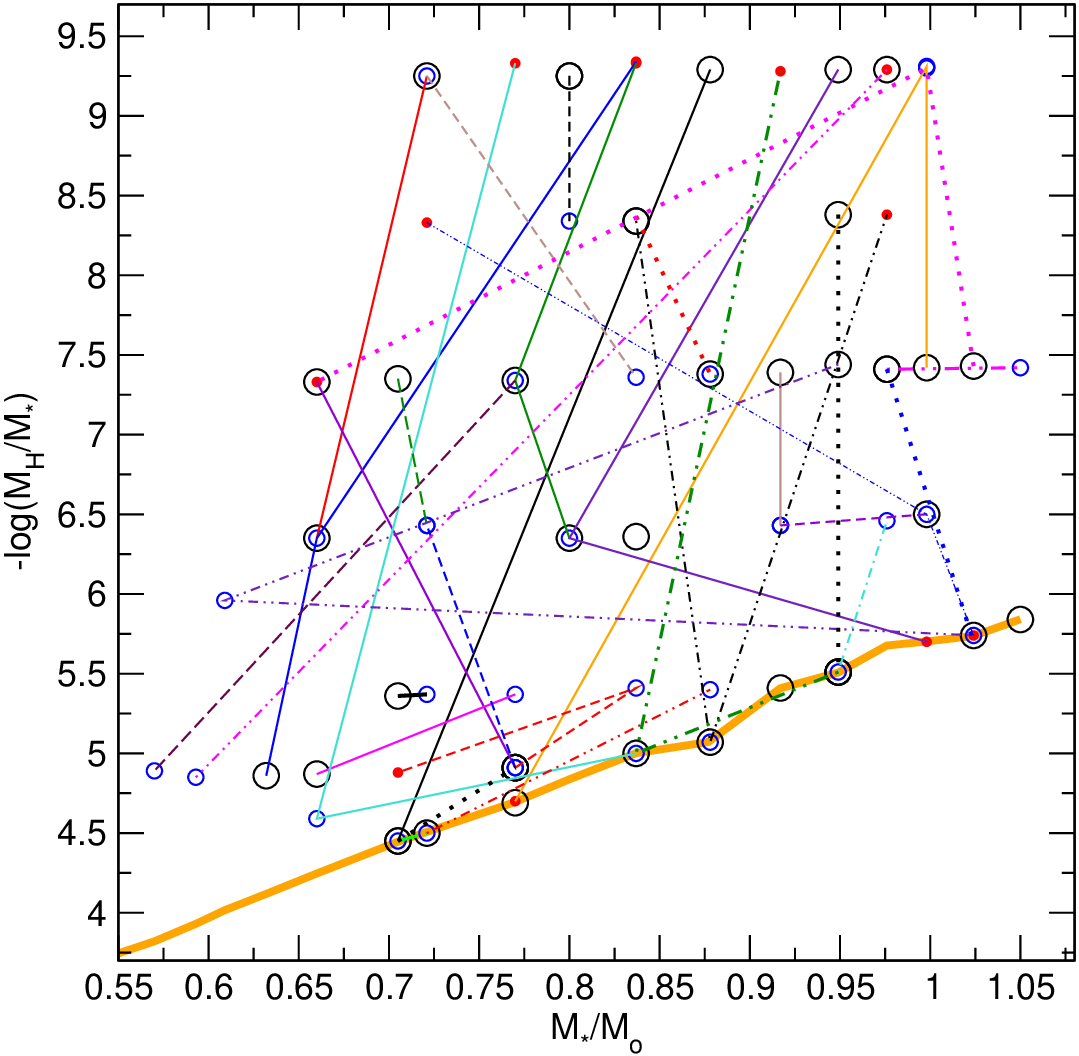}
\caption{The values of the hydrogen envelope mass in 
terms of the stellar mass corresponding to all the 
asteroseismological models of the 42 massive ZZ Ceti 
stars studied in this work. Large black, medium blue and 
little full red circles correspond to the first, second 
and third asteroseismological solution, respectively, for 
each star (see Tables \ref{tabla-periodos-1} and  \ref{tabla-periodos-2}). 
Solutions corresponding to 
the same object are joined together. The  thick orange line  
depicts the canonical  values of the hydrogen envelope 
thickness (color version of this figure is only available 
in the on-line version). }
\label{env-sismology}
\end{center}
\end{figure}

\begin{figure}
\begin{center}
\includegraphics[clip,width=250pt]{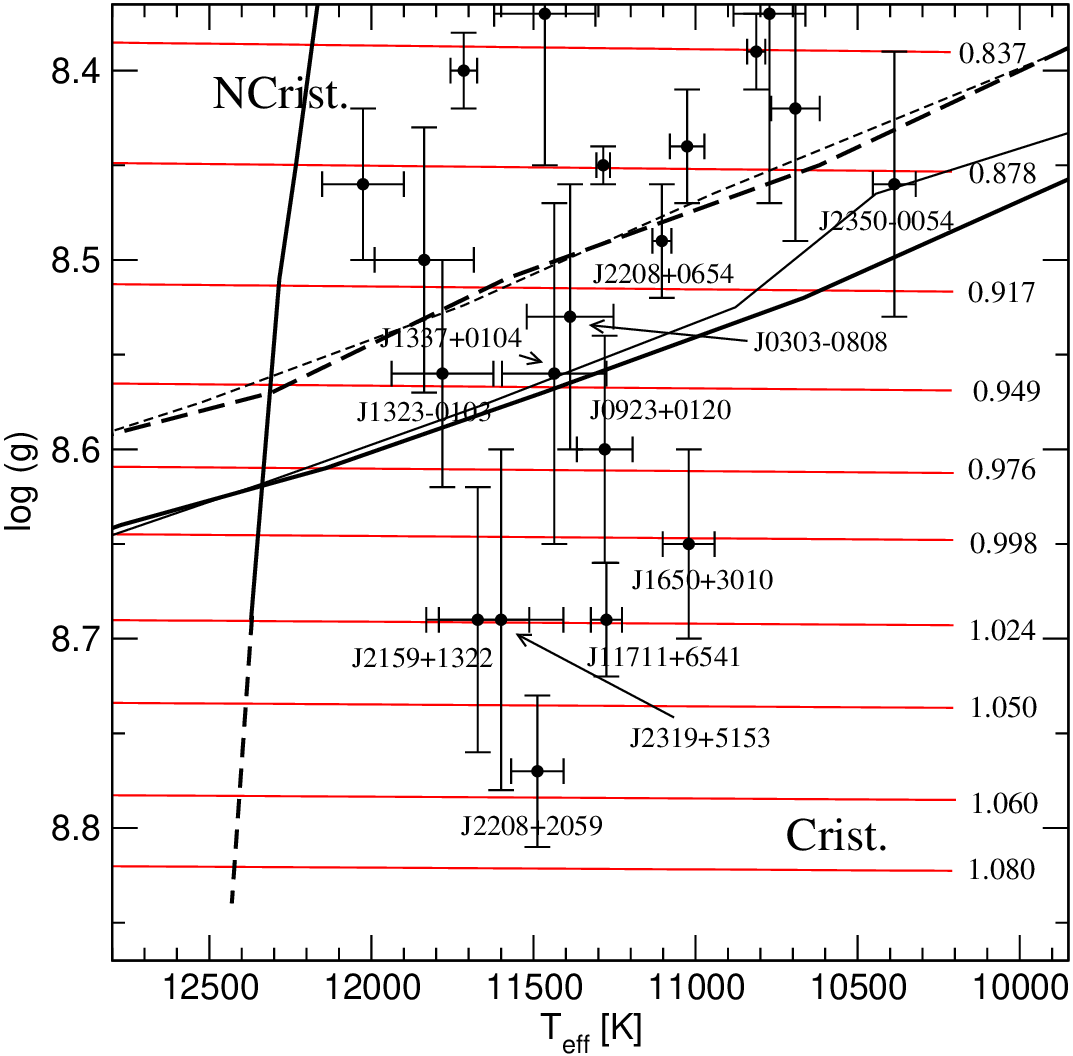}
\caption{A zoom of Figure \ref{gteff} on the region were crystallized DA white 
dwarf should be. The evolutionary tracks employed in our study are shown as red 
lines along with its value of the stellar mass. The thick full (dashed) lines 
indicate the limit between crystallization (Crist.) and no$-$crystallization (NCrist.) for the H2010 
(SC1993) phase diagram corresponding to models with canonical H envelopes, and 
thin lines are for models with the thinner H envelope of each sequence.}
\label{crist-spec}
\end{center}
\end{figure}

In Figure  \ref{env-sismology} we plot  the thickness of  the hydrogen
envelope  in terms  of the  stellar mass  for  the asteroseismological
models listed in Tables \ref{tabla-periodos-1} and \ref{tabla-periodos-2}. 
With black large circles
we plot  the best fit  models for each  star, whereas blue  medium and
small full red circles represent  the second and third solutions, when
present.   Solutions  corresponding  to  the  same  object  are  joint
together with a line.  The gray thick line indicates the high limit of
the hydrogen mass,  as predicted by stellar evolution.   Note that for
several objects,  we obtain two possible  seismological solutions, one
characterized by a high stellar  mass and a thin hydrogen envelope and
other characterized by a lower  mass and a thicker hydrogen layer. For
example, for  J2128$-$0007 we obtained a best  fit model characterized
by $M_*  = 0.976  M_{\odot}$ and $\log(M_{\rm  H}/M_*) = -9.29$  and a
second solution with $M_*  = 0.593 M_{\odot}$ and $\log(M_{\rm H}/M_*)
= -4.85$.   This degeneracy in solutions  is related to  the so called
``core $-$  envelope symmetry'' discussed  in Montgomery et  al. (2003),
where a  sharp feature in  the Brunt$-$V\"ais\"al\"a frequency  in the
envelope can produce  the same period changes as a  bump placed in the
core.

The mean value of the hydrogen layer mass is $\langle M_{\rm H}/M_*) =
5.24  \times 10^{-6}$  for our  sample of  42 massive  DAV  stars, and
$\langle  M_{\rm H}/M_*)  = 4.54  \times 10^{-6}$  if we  consider the
reduced     sample     of     18     stars     listed     in     Table
\ref{tabla-periodos-1}. These  values are about 4 times  lower than the
mean value obtained  by Romero et al. (2012),  with a different sample
of stars but the same model  grid, and about 10 times larger than that
from Castanheira \& Kepler (2009), with a sample with a broad range in
stellar  mass, including  very massive  ZZ Ceti  stars,  and employing
different models. Notwithstanding these differences, our results agree
with those obtained by Castanheira  \& Kepler (2009) and Romero et al.
(2012) in that the possible values of the hydrogen mass are not around
$10^{-4} M_*$ but  span over a large range  ($10^{.4} - 10^{-10} M_*$)
and that an important fraction of DA white dwarf stars might be formed
with an hydrogen envelope much thinner than that predicted by standard
evolutionary theory. This result should have a strong impact on the derived ages from white dwarf cooling sequences for globular clusters, since it is always assumed that the amount of hydrogen in the envelope is $\sim 10^{-4} M_*$.

As  we mentioned  earlier, there  are observational  evidence  for the
existence of a range in the hydrogen layer mass.  Tremblay \& Bergeron
(2008)  determined the ratio  of helium  to hydrogen  atmosphere white
dwarf stars in terms of $T_{\rm eff}$ from a model atmosphere analysis
of the  infrared photometric data from  the Two Micron  All Sky Survey
combined with  available visual  magnitudes. These authors  found that
the  He/H atmosphere ratio  increases gradually  from $\sim  0.25$ for
$15\, 000 K \gtrsim T_{\rm eff} \gtrsim 10\, 000 K$ to $\sim 0.50$ for
$10\, 000 K \gtrsim T_{\rm eff}  \gtrsim 8\, 000 K$, due to convective
mixing when  the bottom  of the hydrogen  convection zone  reaches the
underlying convective  He envelope. They  conclude that about  15\% of
the DA white dwarf should have hydrogen mass layers in the range $\log
(M_{\rm H}/M_*) = -10$ to $-8$.   Romero et al. (2012), based on a set
of 44  bright ZZ  Ceti stars with  stellar mass $\sim  0.6 M_{\odot}$,
found that  $\sim 11 \%$ of  the sample have a  thin hydrogen envelope
mass  in the  range  $10^{-10} \lesssim  M_{\rm H}/M_{\odot}  \lesssim
10^{-8}$. From our asteroseismological results, we found that 7 out of
42 objects in  our sample, $\sim 17 \%$,  have thin hydrogen envelopes
in this range, compatible with the predictions of Tremblay \& Bergeron
(2008).

Recently, Kurtz et al. (2013) reported the discovery of a new class of
pulsating white  dwarf star, the  ``hot DAV'' stars,  characterized by
hydrogen atmospheres and effective  temperatures in the cooler edge of
the  ``DB  gap'',  located  between  $45\,  000  -  30\,  000$ K.   The
pulsational instability of  the ``hot DAV'' white dwarfs  was predicted by
Shibahashi (2005, 2007).  This author found that in  models with thin
hydrogen envelopes and temperatures  around $30\, 000$ K, the radiative
heat  exchange leads to  an asymmetry  in $g-$mode  oscillatory motion
such  that  the   oscillating  elements  overshoot  their  equilibrium
positions with increasing velocity  (Kurtz et al. 2008).  He predicted
that high harmonic  degree $g$-modes should be excited  in DA stars at
the cool edge of the DB  gap.  Since the existence of these objects in
the DB  gap {\it{requires}} that  a fraction of  the ``hot DAV''  stars to
have a thin layer of hydrogen of $\sim 10^{-12} M_{\odot}$ on top of a
much thicker  helium layer, the  discovery of the  ``hot DAV'' stars  is a
{\it{confirmation}}  of  the  existence  of  extremely  thin  hydrogen
envelope DA white dwarf stars.

From our set of evolutionary  sequences we found that the thickness of
the  hydrogen envelope decreases  as the  stars evolves  towards lower
effective  temperatures,  until  it   reaches  a  stable  value.   For
sequences  with hydrogen  layer mass  below $\log  (M_{\rm H}/M_*)\sim
-8$, this usually happens before the  star reaches the cold end of the
DB  gap, so  the thickness  of  the hydrogen  layer at  $30\, 000$  is
practically the  same than that  in the instability  strip. Therefore,
the observed ZZ Cetis must  have hydrogen envelopes thicker than $\sim
10^{-12}M_*$ in order to remain DA white dwarfs when they reach the DA
instability  strip.  Additional computations  performed for  this work
show  that, for  a sequence  with  $0.593 M_{\odot}$  and an  hydrogen
envelope  mass  of  $1.38\times  10^{-10}  M_*$,  the  outer  hydrogen
convective zone reaches the helium rich layer underneath for effective
temperatures around $\sim  9300$ K, mixing the hydrogen  with the much
more abundant helium so the star becomes a DB white dwarf.

\subsection{Crystallization from asteroseismology} 

In Figure \ref{crist-spec} we show a  zoom of the high $\log g$ region
of the $T_{\rm eff} - \log g$ plane from Figure \ref{gteff}.  Diagonal
lines   show   the  limit   between   crystallization  (Crist.)    and
no$-$crystallization  (NCrist.)  considering  the phase  diagrams from
H2010  (solid)  and  SC1993   (dashed).   Thick  lines  correspond  to
sequences    with    canonical    hydrogen   envelope    (see    Table
\ref{teff-crist}),   while   thin    lines   depict   the   onset   of
crystallization for  sequences with the thinnest  hydrogen envelope of
the  grid, for  each  stellar mass.   According  to the  spectroscopic
values of $T_{\rm  eff}$ and $\log g$, there are  $\sim 11$ objects in
our sample that  should have a fraction of its  mass in a crystallized
state according to  the SC1993 phase diagram, and  6 of them according
to H2010 phase diagram. These number increase if we consider the error
bars.  Note  that the  crystallized mass fraction  for a  given object
will differ, depending on which  phase diagram we are employing in our
computations, being usually $\sim 10\%$  lower for H2010.  It is worth
noting that the crystallization  temperatures when only the release of
latent heat  is taken  into account, are  similar to those  for SC1993
phase diagram.

Our     asteroseismological      results,     listed     in     Tables
\ref{tabla-periodos-1} and  \ref{tabla-periodos-2}, show that  15 out
of the 42 massive DAV stars analyzed in this work have best fit models
corresponding  to a sequence  computed by  considering the  release of
gravitational energy due to phase separation upon crystallization, and
consequently  showing some  fraction  of its  mass  in a  crystallized
state.  For nine of them, the best fit model corresponds to a sequence
computed by  using the  H2010 phase diagram,  while for  the remaining
six stars we obtain a best fit model corresponding to sequences where
we consider  the SC1993  phase diagram.  Note  that from the  15 stars
with  crystallized  models, nine  also  show  three  or more  observed
periods   (see  Table  \ref{tabla-periodos-1}),   six  of   them  with
seismological solutions corresponding  to sequences computed using the
H2010  phase  diagram  and   three  with  the  SC1993  phase  diagram.
Therefore, our results from asteroseismology suggest that the Horowitz
et   al.   (2010)   phase  diagram   should  be   the   most  accurate
representation  of  the  crystallization  process inside  white  dwarf
stars.  We  must admit  that in  most cases we  present more  than one
possible  solution for  a single  object,  with quality
functions values  that are  quite close in  some cases.   In addition,
most of the stars analyzed in this work show only a few, and sometimes
one,  modes in their  observed spectrum,  that limits  the information
that  can be  extracted from  asteroseismology. However,  in  spite of
these, we  must recall that a  set of rigorous criteria  is applied in
order to  select for each  star in our sample  the asteroseismological
model that better  matches not only the observed  periods but also the
spectroscopic  parameters and  any additional  observational  data. In
this sense we consider that our results are robust, although it should
be necessary to analyze a large number of objects to place our finding
on a more firmer basis.

Finally, note  that the carbon$-$oxygen abundances  are not considered
as free parameters in our model grid. Instead, the chemical abundances
of the central regions are basically given by the evolution during the
central     helium    burning    stage,     where    we     use    the
$^{12}$C($\alpha,\gamma$)$^{16}$O reaction rate given by Angulo et al.
(1999) in our computations.  However, we must recall that the specific
value of the $^{12}$C($\alpha, \gamma$)$^{16}$O reaction rate is still
one of the current uncertainties in the theory of stellar evolution. A
higher $^{12}$C($\alpha, \gamma$)$^{16}$O reaction rate will translate
into  a  higher   oxygen  abundance,  increasing  the  crystallization
temperature of the model.  Our additional computations show that if we
increase  by  a factor  of  1.4 the  $^{12}$C($\alpha,\gamma$)$^{16}$O
reaction rate from Angulo et al. (1999) the amount of oxygen left in a
$0.998M_{\odot}$ white  dwarf model will  increase by $\sim  14.5 \%$,
and  the  crystallization  temperature  given  by  the  phase  diagram
presented by  Horowitz et al. (2010)  increases by about  $\sim 700$ K
from   that   listed   in   Table  \ref{teff-crist}.    Thus,   higher
crystallization  temperatures  can   be  achieved  by  increasing  the
$^{12}$C($\alpha,\gamma$)$^{16}$O    reaction    rate,   within    its
uncertainty.

\section{Conclusions}
\label{conclusions}
In this paper we have  carried out the first asteroseismological study
applied to  massive variable DA  white dwarf stars supposed  to harbor
carbon$-$oxygen  cores.   To  this  end  we  employ  a  set  of  fully
evolutionary models  characterized by detailed  chemical profiles from
the  center to  the surface.  We  study a  sample of  42 objects  with
spectroscopic stellar mass in the  range of $0.72 - 1.05 M_{\odot}$. A
first version of our model grid  was employed by Romero et al.  (2012)
in an  asteroseismological study of  44 bright ZZ Ceti,  including the
most  stable G117$-$B15A. In  this work  we extend  our model  grid to
higher stellar masses  to achieve a full coverage  of the stellar mass
range where massive DA stars  are found.  In addition, we introduce an
additional parameter  given by the  crystallization treatment. Besides
the computations where we only  include the release of latent heat, we
compute  evolutionary  sequences   by  employing  the  phase  diagrams
presented in Horowitz et al.  (2010) and Segretain \& Cheabier (1993),
accounting  for the  release of  energy due  to phase  separation upon
crystallization.

Our main results from our asteroseismological study are the following:

\begin{itemize}
 \item  We  redetermine  the  spectroscopic  mass of all the  42  objects
    using  our DA  white dwarf evolutionary  tracks, including
   sequences  with  stellar  masses  higher than  1.05$M_{\odot}$,  with
   oxygen$-$neon cores, in order  to achieve a better determination in
   the higher limit of our sample.
   We obtain a mean spectroscopic mass of $\langle M_* \rangle_{\rm
  spec} =  0.841 \pm  0.093 M_{\odot}$.  From  our asteroseismological
  fits,  we   obtain  a  mean  seismological  mass   of  $\langle  M_*
  \rangle_{\rm  seis} =  0.850 \pm  0.110 M_{\odot}$,  slightly higher
  than the  spectroscopic value. Since both values  are obtained based
  on two different techniques we consider that the agreement is excellent.

\item In  line with previous  asteroseismological studies (Castanheira
  \& Kepler 2009; Romero et al. 2012), we find a range of thickness of
  the  hydrogen   envelope.  The
  distribution is not  homogeneous, but shows three peaks.  One peak for
  thick  envelopes at  $\log(M_{\rm  H})\sim  -5.5$  mostly
  composed  by  models  with  canonical  envelopes,  as  predicted  by
  evolutionary computations, a  second strong peak around $\log(M_{\rm
    H})  \sim -7.5$,  and  a weak  third  peak at  very thin  envelopes
  $\log(M_{\rm H}) \sim -9.5$.

\item We find that 32 out of the 42 ZZ Ceti stars analyzed have a best
  fit  model  characterized  by  an  hydrogen  envelope  thinner  than
  predicted by  standard evolutionary  theory, with a  strong component
  between $\log(M_{\rm H}) = -7$ and $-8$.

\item  We  find a  mean  hydrogen  envelope  mass of  $\langle  M_{\rm
  H}/M_*\rangle = 5.24  \times 10^{-6}$, 4 times lower  than the value
  obtained by Romero et al. (2012).  This difference may be due to the
  stellar mass  dependence of  the hydrogen mass,  since the  range of
  stellar  masses  of  our sample  is  higher  than  that in  Romero  et
  al. (2012), so the hydrogen  envelope of our sample should be intrinsically
  thinner. Finally, if we consider only those stars having three or more periods
  the mean hydrogen envelope mass is  $4.54  \times 10^{-6} M_*$. 
\end{itemize}

We  also  study the  impact  of  the  crystallization process  on  the
pulsation  spectrum of massive  variable white  dwarfs.  We  find that
crystallization does  affect the pulsation properties  of these stars.
We  find  that  the periods  increases  $\sim  10  -20$ s  when  phase
separation  upon crystallization  is considered  in  the computations.
This increase is  noticeable for modes with periods  larger than $\sim
250$ s,  or radial order  $k \gtrsim 4$.   The period spacing  is also
affected by  the action of crystallization process,  mostly because of
the  changes  in  the  inner  chemical  profile  due  to  the  growing
crystallized core.  As the fraction of crystallized mass increases the
period  spacing  becomes smoother  and  the  modes  are closer  to  an
harmonic configuration,  with a $\Delta  \Pi$ close to  its asymptotic
value.

From our asteroseismological fits, we find that 15 stars have best fit
models showing a fraction of its mass in a crystallized state. Nine of
them are  best fitted by sequences  where the Horowitz  et al.  (2010)
phase  diagram was  employed.  The  remaining six  are best  fitted by
sequences  characterized  by  a  Segretain \&  Chabrier  (1993)  phase
diagram. Then, our asteroseismological results indicate that the phase
diagram presented  in Horowitz et al.   (2010) is the  one that better
represent  the  crystallization process  in  white  dwarf stars.   Our
result is in agreement with the  results of Winget et al. (2009, 2010)
based on the study of  the white dwarf luminosity function in globular
clusters.

\section*{Acknowledgments}

We thank an anonymous referee for important comments and suggestions.
Part of this work was supported by  CNPq-Brazil and AGENCIA through  the
Programa de  Modernizaci\'on Tecnol\'ogica BID 1728/OC-AR,  by the PIP
112-200801-00940 grant  from CONICET.  Part of this work  has been done  
with observations from the Southern Astrophysical Research (SOAR) telescope,
which  is  a  joint  project  of the  Minist\'{e}rio  da  Ci\^{e}ncia,
Tecnologia, e  Inova\c{c}\~{a}o (MCTI) da  Rep\'{u}blica Federativa do
Brasil, the  U.S. National  Optical Astronomy Observatory  (NOAO), the
University of North Carolina at  Chapel Hill (UNC), and Michigan State
University (MSU).   This research has made use  of NASA's Astrophysics
Data System.

\newpage

\begin{table}
\centering
\caption{The main characteristics of our set of DA white dwarf models. The values 
of the stellar mass are listed in column 1. 
The hydrogen mass corresponding to standard 
evolutionary computations is listed, for each mass, in column 2, along with 
the helium mass (column 3). Columns 4 and 5 show the central abundances of 
carbon and oxygen for each sequence. }
\begin{tabular}{ccccc}
\hline\hline
$M_* / M_{\odot}$ & $-\log (M_{\rm H} / M_*)$ & $-\log (M_{\rm He} / M_*)$ & $X_{\rm C}$ & $X_{\rm O}$ \\
\hline\hline
0.525 & 3.62 & 1.31 & 0.278 & 0.709 \\
0.548 & 3.74 & 1.38 & 0.290 & 0.697 \\
0.570 & 3.82 & 1.46 & 0.301 & 0.696 \\
0.593 & 3.93 & 1.62 & 0.283 & 0.704 \\
0.609 & 4.02 & 1.61 & 0.264 & 0.723 \\ 
0.632 & 4.25 & 1.76 & 0.234 & 0.755 \\
0.660 & 4.26 & 1.92 & 0.258 & 0.730 \\
0.705 & 4.45 & 2.12 & 0.326 & 0.661 \\
0.721 & 4.50 & 2.14 & 0.328 & 0.659 \\
0.770 & 4.70 & 2.23 & 0.332 & 0.655 \\
0.800 & 4.84 & 2.33 & 0.339 & 0.648 \\
0.837 & 5.00 & 2.50 & 0.347 & 0.640 \\
0.878 & 5.07 & 2.59 & 0.367 & 0.611 \\
0.917 & 5.41 & 2.88 & 0.378 & 0.609 \\
0.949 & 5.51 & 2.92 & 0.373 & 0.614 \\
0.976 & 5.68 & 2.96 & 0.374 & 0.613 \\
0.998 & 5.70 & 3.11 & 0.358 & 0.629 \\
1.024 & 5.74 & 3.25 & 0.356 & 0.631 \\
1.050 & 5.84 & 2.96 & 0.374 & 0.613 \\
\hline\hline
\label{tabla-grid}
\end{tabular}
\end{table}

\begin{table}
\centering
\caption{Effective temperature and surface gravity when crystallization 
process starts for a given stellar mass, considering Horowitz et al. (2010)  
 (columns 2 and 3) and Segretain \& Chabrier (1993) (columns 4 and 5) phase diagrams. 
These values correspond to the sequences with thick hydrogen envelopes, as predicted 
by standard stellar evolution.}
\begin{tabular}{crcrc}
\hline
 & \multicolumn{2}{c}{Horowitz et al. (2010)} & \multicolumn{2}{c}{Segretain \& Chabrier (1993)} \\
\hline
$ M_* / M_{\odot}$ &   $T_{\rm eff}$ [K] &    $\log g$    &  $T_{\rm eff}$ [K] &    $\log g$  \\
\hline
0.800  &  8401 & 8.34 &   9150  & 8.33 \\
0.837  &  9053 & 8.39 &   9874  & 8.39 \\
0.878  &  9753 & 8.45 &  10618  & 8.45 \\ 
0.917  & 10666 & 8.52 &  11585  & 8.51 \\ 
0.949  & 11470 & 8.57 &  12308  & 8.57 \\
0.976  & 12142 & 8.61 &  13203  & 8.61 \\
0.998  & 12773 & 8.64 &  13802  & 8.64 \\ 
1.024  & 13594 & 8.69 &  14664  & 8.69 \\ 
1.050  & 14462 & 8.73 &  15611  & 8.73 \\
\hline
\label{teff-crist}
\end{tabular}
\end{table}

\begin{table}
\caption{Spectroscopic parameters and the derived spectroscopic mass for the massive 
ZZ Ceti stars of our sample.}
\centering
\scalebox{0.8}[0.8]{ \hspace{-7mm}
\begin{tabular}{lcccc}
\hline\hline
 ${\rm Star}$  &   $T_{\rm eff}$ [K] &    $\log g$    & $M_*/M_{\odot}$   &  Ref.\\
\hline\hline
 J0000$-$0046 &  $10\,772 \pm  111$  & $8.37\pm 0.10$  & $0.825\pm 0.063$   &   1\\ 
 J0048$+$1521 &  $11\,260 \pm  131$  & $8.33\pm 0.07$  & $0.801\pm 0.044$   &   1\\
 J0102$-$0032 &  $11\,024 \pm  106$  & $8.26\pm 0.08$  & $0.756\pm 0.050$   &   1\\
 J0111$+$0018 &  $11\,765 \pm   91$  & $8.32\pm 0.04$  & $0.795\pm 0.025$   &   1\\
 J0249$-$0100 &  $11\,070 \pm  129$  & $8.24\pm 0.10$  & $0.743\pm 0.068$   &   1\\
 J0303$-$0808 &  $11\,387 \pm  134$  & $8.53\pm 0.07$  & $0.926\pm 0.043$   &   1\\
 J0322$-$0049 &  $11\,040 \pm   70$  & $8.25\pm 0.06$  & $0.749\pm 0.038$   &   1\\
 J0349$+$1036 &  $11\,715 \pm   41$  & $8.40\pm 0.02$  & $0.845\pm 0.012$   &   1\\
 J0825$+$0329 &  $11\,969 \pm  117$  & $8.25\pm 0.04$  & $0.751\pm 0.025$   &   1\\
 J0825$+$4119 &  $11\,837 \pm  153$  & $8.50\pm 0.07$  & $0.901\pm 0.044$   &   1\\
 J0843$+$0431 &  $11\,268 \pm   71$  & $8.22\pm 0.04$  & $0.732\pm 0.025$   &   1\\
 J0855$+$0635 &  $11\,026 \pm   53$  & $8.44\pm 0.03$  & $0.870\pm 0.019$   &   1\\
 J0923$+$0120 &  $11\,280 \pm   86$  & $8.60\pm 0.06$  & $0.969\pm 0.037$   &   1\\
 J0925$+$0509 &  $10\,813 \pm   28$  & $8.39\pm 0.02$  & $0.838\pm 0.013$   &   1\\
 J0939$+$5609 &  $11\,790 \pm  160$  & $8.22\pm 0.07$  & $0.732\pm 0.047$   &   1\\
 J0940$-$0052 &  $10\,692 \pm   75$  & $8.42\pm 0.07$  & $0.856\pm 0.044$   &   1\\ 
 J1105$-$1613 &  $11\,677 \pm   87$  & $8.23\pm 0.03$  & $0.738\pm 0.018$   &   1\\
 J1200$-$0251 &  $11\,986 \pm  143$  & $8.33\pm 0.06$  & $0.802\pm 0.038$   &   1\\
 J1216$+$0922 &  $11\,344 \pm  125$  & $8.30\pm 0.07$  & $0.782\pm 0.044$   &   1\\
 J1218$+$0042 &  $11\,060 \pm   80$  & $8.21\pm 0.06$  & $0.725\pm 0.042$   &   1\\
 J1222$-$0243 &  $11\,421 \pm   52$  & $8.27\pm 0.03$  & $0.763\pm 0.019$   &   1\\
 J1257$+$0124 &  $11\,465 \pm  156$  & $8.37\pm 0.08$  & $0.826\pm 0.051$   &   1\\
 J1323$+$0103 &  $11\,781 \pm  157$  & $8.56\pm 0.06$  & $0.945\pm 0.037$   &   1\\
 J1337$+$0104 &  $1\,1436 \pm  161$  & $8.56\pm 0.09$  & $0.945\pm 0.056$   &   1\\
 J1612$+$0830 &  $12\,026 \pm  126$  & $8.46\pm 0.04$  & $0.884\pm 0.026$   &   1\\
 J1641$+$3521 &  $11\,306 \pm  185$  & $8.27\pm 0.11$  & $0.763\pm 0.070$   &   1\\
 J1650$+$3010 &  $11\,021 \pm   80$  & $8.65\pm 0.05$  & $0.999\pm 0.030$   &   1\\
 J1711$+$6541 &  $11\,275 \pm   48$  & $8.69\pm 0.03$  & $1.023\pm 0.018$   &   1\\
 J2128$-$0007 &  $11\,395 \pm  106$  & $8.31\pm 0.07$  & $0.788\pm 0.044$   &   1\\
 J2159$+$1322 &  $11\,672 \pm  159$  & $8.69\pm 0.07$  & $1.023\pm 0.041$   &   1\\
 J2208$+$0654 &  $11\,104 \pm   29$  & $8.49\pm 0.03$  & $0.901\pm 0.019$   &   1\\
 J2208$+$2059 &  $11\,488 \pm   81$  & $8.77\pm 0.04$  & $1.057\pm 0.017$   &   1\\ 
 J2209$-$0919 &  $11\,756 \pm  148$  & $8.34\pm 0.06$  & $0.807\pm 0.038$   &   1\\
 J2214$-$0025 &  $11\,560 \pm   95$  & $8.32\pm 0.05$  & $0.795\pm 0.031$   &   1\\
 J2319$+$5153 &  $11\,600 \pm  192$  & $8.69\pm 0.09$  & $1.023\pm 0.053$   &   1\\  
 J2350$-$0054 &  $10\,387 \pm   66$  & $8.46\pm 0.07$  & $0.882\pm 0.044$   &   1\\
 J1916$+$3938 &  $11\,129 \pm  115$  & $8.34\pm 0.06$  & $0.805\pm 0.040$   &   2\\
 G226$-$29    &  $12\,260 \pm  300$  & $8.31\pm 0.12$  & $0.789\pm 0.075$   &   3,4,5*\\
 L19$-$2      &  $12\,100 \pm  200$  & $8.21\pm 0.10$  & $0.727\pm 0.062$   &   3,4*\\  
 G207$-$9     &  $11\,950 \pm  200$  & $8.35\pm 0.10$  & $0.814\pm 0.064$   &   3,4*\\
 EC0532$-$560 &  $11\,285 \pm   21$  & $8.45\pm 0.01$  & $0.877\pm 0.068$   &   6 \\
 BPM30551     &  $11\,260 \pm  200$  & $8.23\pm 0.05$  & $0.738\pm 0.032$   &   3*\\ 
\hline\hline
\end{tabular}
\tablerefs{ References: (1) Kleinman et al. (2013),  (2) Hermes et al. (2011), (3) Bergeron et al. (2004), (4) Koester \& Allard (2000), (5) Gianninas et al. (2005),  (6) Koester et al. (2009)}
\tablecomments{* Analyzed in Romero et al. (2012)}}
\label{tabla-masas}
\end{table}

\begin{table}
\centering
\caption{Journal of observations for the ZZ Ceti stars reobserved using the 4.1 m SOAR telescope.}
\begin{tabular}{l|c|c}
\hline\hline
Star & Date of Obs. & Length (h) \\
\hline\hline
SDSS J092511.61+050932.44 & 2010-04-13 & 2.0\\
                          & 2010-04-14 & 3.0\\
L19-2                     & 2011-06-15 & 3.2\\
SDSS J132350.28+010304.22 & 2012-06-13 & 3.1\\
                          & 2012-06-14 & 3.1\\
SDSS J122229.58-024332.54 & 2013-05-13 & 4.0\\
SDSS J125710.50+012422.89 & 2013-05-13 & 4.0\\
\hline\hline
\end{tabular}
\label{obs}
\end{table}

\begin{deluxetable}{cccccccccccc}
\tabletypesize{\tiny}
\tablecaption{Results from asteroseismological fits for the 18 massive ZZ Ceti stars with carbon$-$oxygen core that show three or more observed periods in their spectrum. For each star we list the main structure parameters from the seismological solutions, along with the observed and theoretical periods and the corresponding value of the quality function $\Phi$. The values of the harmonic degree $\ell$  and radial order $k$ are listed for each theoretical period. The observed periods and amplitudes are extracted from different works listed in the last column (see text for details).}
\tablewidth{0pt}
\tablehead{
\colhead{${\rm star}$} & \colhead{$T_{\rm eff}$} & \colhead{$M_*/M_{\odot}$} & \colhead{$\log(M_{\rm H}/M_*)$} & \colhead{$\Pi_i^{\rm obs}$} &  \colhead{A}    & \colhead{$\Pi_k^{\rm th}$} & \colhead{$\ell$} & \colhead{$k$} & \colhead{$\Phi$} & \colhead{Crist.} & \colhead{Ref.}\\
    & $[$K$]$     &        &    &   $[$s$]$      & (mma) & $[$s$]$       &     &     &     & \% & }  
\startdata
 J0000$-$0046 & 11352 & 0.878 & -9.29 & 584.82 & 15.9 & 584.679 & 2 & 21 & 0.541 &  & Castanheira et al. (2006)\\
              &       &       &       & 601.35 & 8.97 & 601.493 & 2 & 22 &   & & \\
              &       &       &       & 611.42 & 23.0 & 610.278 & 1 & 12 & &  & \\
              & 11417 & 0.705 & -4.45 & 584.82 & 15.9 & 584.342 & 1 & 12 & 0.5776 &  &  \\
              &       &       &       & 601.35 & 8.97 & 601.440 & 2 & 23 && & \\
              &       &       &       & 611.42 & 23.0 & 610.369 & 1 & 13 & & &\\
\hline
 J0048$+$1521 & 11470 & 0.949 & -5.51 & 323.14 & 14.75 & 323.052 & 2 & 14 & 1.983 & H NCrist. & Mullally et al. (2005)\\
              &       &       &       & 333.18 & 8.58  & 333.448 & 1 & 7  & && \\
              &       &       &       & 609.75 & 22.10 & 606.328 & 1 & 16 & & & \\
              &       &       &       & 636.41 & 8.71  & 636.586 & 2 & 30 && &  \\
              &       &       &       & 672.30 & 8.49  & 673.282 & 1 & 18 && &\\
              &       &       &       & 698.36 & 18.29 & 705.145 & 1 & 19 && & \\   
\hline
 J0825$+$0329 & 11419 & 0.770 & -5.37 & 640.23 & 5.18 & 640.319 & 2 & 25 & 0.936 & &Kepler et al. (2005)\\
              &       &       &       & 657.36 & 4.42 & 655.806 & 1 & 14 &&&\\ 
              &       &       &       & 704.11 & 3.74 & 705.206 & 1 & 15 &&& \\     
              & 11406 & 0.660 & -4.87 & 640.23 & 5.18 & 639.955 & 2 & 23 & 0.618 &&\\
              &       &       &       & 657.36 & 4.42 & 658.140 & 1 & 13 & &&\\
              &       &       &       & 704.11 & 3.74 & 704.699 & 1 & 14 &&&\\
\hline
 J1200$-$0251 & 11715 & 0.917 & -6.43 & 257.10 & 6.69  & 256.478 & 1 & 4 & 1.714 &  H (Ncrist.) & Castanheira et al. (2013a) \\
              &       &       &       & 271.30 & 13.09 & 274.495 & 2 & 10 &&&\\
              &       &       &       & 294.10 & 6.69  & 294.399 & 1 & 5 &&&\\
              &       &       &       & 304.78 & 23.72 & 304.506 & 2 & 12 &&&\\      
              & 12180 & 0.998 & -6.50 & 257.10 & 6.69  & 253.244 & 1 & 4 & 1.135 &  SC 21.34 & \\
              &       &       &       & 271.30 & 13.09 & 271.340 & 2 & 11 &&&  \\
              &       &       &       & 294.10 & 6.69  &  293.150 & 2 & 12 &&&\\
              &       &       &       & 304.78 & 23.72 & 305.308 & 1 & 6 & &&\\   
\hline
 J1216$+$0922 & 11658 & 0.770 & -7.34 & 409 & 30.1 & 409.092 & 1 & 6 & 1.671 & & Kepler et al. (2005)\\
              &       &       &       & 570 & 24.6 & 568.389 & 2 & 20 & & &Castanheira \& Kepler (2009)\\
              &       &       &       & 626 & 21.6 & 625.764 & 1 & 12 &&&\\
              &       &       &       & 823 & 45.2 & 824.526 & 2 & 30 &&&\\ 
              &       &       &       & 840 & 42.0 & 838.389 & 1 & 17 &&&\\
              &       &       &       & 967 & 20.5 & 970.565 & 1 & 20 &&&\\
              & 11554 & 0.570 & -4.89 & 409 & 30.1 & 407.208 & 2 & 12 & 1.395 &&\\
              &       &       &       & 570 & 24.6 & 567.609 & 1 & 9 &&&\\
              &       &       &       & 626 & 21.6 & 625.430 & 2 & 20 &&&\\
              &       &       &       & 823 & 45.2 & 822.982 & 1 & 15 &&&\\
              &       &       &       & 840 & 42.0 & 840.050 & 2 & 28 &&&\\
              &       &       &       & 967 & 20.5 & 968.756 & 1 & 18 &&&\\
\hline
 J1222$-$0243 & 11180 & 0.837 & -7.36 & 350.371  & 3.393  & 351.761  & 2 & 12 & 1.932  & & This paper\\
              &       &       &       & 395.976  & 18.215 & 395.955  & 1 & 6  &        & &\\
              &       &       &       & 842.173  & 2.716  & 843.878  & 2 & 32 &        & &\\
              &       &       &       & 1177.309 & 3.834  & 1175.985 & 2 & 45 &        & &\\
              & 11396 & 0.721 & -9.25 & 350.371  & 3.393  & 353.032  & 1 & 4  & 1.352  & &\\
              &       &       &       & 395.976  & 18.215 & 394.044  & 1 & 5  &        & &\\
              &       &       &       & 842.173  & 2.716  & 841.962  & 2 & 22 &        & &\\
              &       &       &       & 1177.309 & 3.834  & 1177.751 & 1 & 27 &        & &\\
 \hline
 J1257$+$0124 & 11172 & 0.705 & -4.45 & 377.838 & 6.607 & 377.792 & 2  & 13 & 3.005 & & This paper\\
              &       &       &       & 398.056 & 6.707 & 400.262 & 1  &  7 &  &&\\
              &       &       &       & 466.274 & 8.941 & 461.018 & 2  & 17 &  && \\
              &       &       &       & 507.060 & 8.404 & 506.337 & 2  & 19 &  &&\\
              &       &       &       & 644.501 & 21.944 & 658.082 & 1 & 14 &  &&\\
              &       &       &       & 786.884 & 8.909 & 785.373 & 1 & 17 &  & &\\
              &       &       &       & 946.257 & 10.451 & 947.977 & 1 & 21 &  && \\
              &       &       &       & 1070.455 & 7.903 & 1068.627 & 1 & 24 &  && \\   
              & 11287 & 0.976 & -7.41 & 377.838 & 6.607 & 378.074 & 1 & 7 & 3.370 & H 14.453& \\
              &       &       &       & 398.056 & 6.707 & 411.607 & 1 & 8 &  &&\\
              &       &       &       & 466.274 & 8.941 & 474.110 & 1 & 10 &  && \\
              &       &       &       & 507.060 & 8.404 & 506.212 & 2 & 20 &  &&\\
              &       &       &       & 644.501 & 21.944 & 665.955 & 1 & 15 &  &&\\
              &       &       &       & 786.884 & 8.909 & 786.427 & 2 & 32 &  && \\
              &       &       &       & 946.257 & 10.451 & 945.425 & 1 & 22 &  && \\
              &       &       &       & 1070.455 & 7.903 &  1069.601 & 1 & 25 && \\               
\hline
 J1323$+$0103 & 11535 & 0.917 & -5.41 & 432.483 & 5.13 & 431.038 & 1 & 10 & 2.794 & & Kepler et al. (2012)\\
              &       &       &       & 497.402 & 6.35 & 489.319 & 1 & 12 & & & This paper \\ 
              &       &       &       & 525     & 3.6  & 526.144 & 1 & 13 & & & \\
              &       &       &       & 550.474 & 8.60 & 549.806 & 2 & 25 &&&\\  
              &       &       &       & 564.552 & 18.31 & 566.150 & 2 & 26 &&&\\ 
              &       &       &       & 590.13  & 7.1  & 590.280 & 2 & 27 &&&\\  
              &       &       &       & 603.623 & 8.27 & 600.018 & 1 & 15 &&&\\                                  
              &       &       &       & 612.23  & 11.9 & 613.221 & 2 & 28 & && \\
              &       &       &       & 636.39  & 4.8  & 635.727 & 2 & 29 &&&\\
              &       &       &       & 656.025 & 15.25 & 657.536 & 2 & 30 &&& \\
              &       &       &       & 675.363 & 6.35  & 676.435 & 2 & 31 &&&\\
              &       &       &       & 698.64  & 4.3  & 693.277 & 1 & 18 &&&\\
              &       &       &       & 731.632 & 5.17 & 730.338 & 2 & 34 &&&\\ 
              &       &       &       & 831.06  & 4.6  & 831.918 & 2 & 39 &&&\\
              &       &       &       & 884.17  & 4.1  & 880.046 & 2 & 41 &&&\\
\hline
 J1711$+$6541 & 11280 & 0.976 & -6.46 & 214.3 & 1.7 &  215.168 & 2 & 7 & 3.114 & SC 28.56 & Mukadam et al. (2009)\\
            &       &       &       & 234.0 & 1.2 &  233.238 & 2 & 8 & & & Castanheira \& Kepler (2009) \\
            &       &       &       & 561.5 & 3.0 &  561.030 & 2 & 23 && &  \\
            &       &       &       & 609.5 & 5.5 &  610.004 & 1 & 14 &&&\\
            &       &       &       & 690.2 & 3.3 &  690.971 & 1 & 16 &&&\\
            &       &       &       & 934.8 & 2.9 &  933.899 & 2 & 39 &&&\\
            &       &       &       & 1186.6 & 3.3 & 1191.376 & 2 & 47 &&&\\
            &       &       &       & 1248.2 & 3.2 & 1241.512 & 1 & 30 &&&\\
            & 11418 & 0.949 & -5.51 & 214.3 & 1.7 &  217.413 & 1 & 3 & 3.265 & H 0.28 &\\
            &       &       &       & 234.0 & 1.2 &  231.744 & 1 & 4 &&  & \\
            &       &       &       & 561.5 & 3.0 &  556.284 & 1 & 14 &&&  \\
            &       &       &       & 609.5 & 5.5 &  609.576 & 1 & 16 &&&\\
            &       &       &       & 690.2 & 3.3 &  692.162 & 2 & 33 &&&\\
            &       &       &       & 934.8 & 2.9 &  935.107 & 2 & 45 &&&\\
            &       &       &       & 1186.6 & 3.3 & 1187.254 & 2 & 57 &&&\\
            &       &       &       & 1248.2 & 3.2 & 1250.833 & 2 & 58 &&&\\

\hline
 J2128$-$0007 & 11569 & 0.976 & -9.29 & 274.9 & 11.0 & 274.859 & 2 & 9 & 0.333 & H 9.14 & Castanheira et al. (2006)\\
              &       &       &       & 289.0 & 9.7  & 289.158 & 2 & 10 & & & \\
              &       &       &       & 302.2 & 17.1 & 301.545 & 1 & 5 &&&\\
              & 11999 & 0.593 & -4.85 & 274.9 & 11.0 & 275.925 & 1 & 3 & 0.667 && \\
              &       &       &       & 289.0 & 9.7  & 289.274 & 2 & 8 &&& \\
              &       &       &       & 302.2 & 17.1 & 301.711 & 1 & 4 &&& \\
              & 11792 & 0.976 & -9.29 & 274.9 & 11.0 & 274.786 & 2 & 9  &  0.578 & SC 17.84 &\\
              &       &       &       & 289.0 & 9.7  & 289.581 & 2 & 10 &&&\\
              &       &       &       & 302.2 & 17.1 & 301.684 & 1 & 5  &&& \\
\hline
 J2208$+$2059 & 11355 & 1.050 & -5.84 &  249.72  & 1.536 &  248.722  &  2  &  11  &  2.215  & H 41.39 & Castanheira et al. (2013b)\\
              &       &       &       &  477.35  & 1.609 &  473.525  &  2  &  23  &         & & \\
              &       &       &       &  538.84  & 5.178 &  538.684  &  2  &  26  &         &&\\ 
              &       &       &       &  558.89  & 8.537 &  559.430  &  1  &  15  &         &&\\ 
              &       &       &       &  576.05  & 2.714 &  577.352  &  2  &  28  &         &&\\
              &       &       &       &  592.60  & 4.500 &  593.614  &  2  &  29  &         &&\\
\hline         
 J2209$-$0919 & 11944 & 0.949 & -7.44 & 221.9 & 4.7 & 220.647 & 2 & 7 & 1.653 & SC 6.46  & Castanheira et al. (2007)\\
              &       &       &       & 294.7 & 5.1 & 293.546 & 2 & 11 &&&\\
              &       &       &       & 448.2 & 10.9 & 448.078 & 2 & 18 &&&\\
              &       &       &       & 789.5 & 10.2 & 788.763 & 2 & 33 &&&\\
              &       &       &       & 894.7 & 43.8 & 894.858 & 1 & 21 &&& \\
              &       &       &       & 968.7 & 6.3 & 973.022 & 1 & 23 &&&\\
              & 11378 & 0.609 & -5.96 & 221.9 & 4.7 & 219.543 & 2 & 5 & 1.868 &&\\
              &       &       &       & 294.7 & 5.1 & 295.898 & 1 & 4 &&&\\
              &       &       &       & 448.2 & 10.9 & 450.580 & 2 & 13 &&& \\
              &       &       &       & 789.5 & 10.2 & 788.924 & 1 & 14 &&& \\
              &       &       &       & 894.7 & 43.8 & 894.099 & 1 & 15 &&& \\
              &       &       &       & 968.7 & 6.3 & 968.758 & 2 & 32 &&&\\
              & 11953 & 1.024 & -5.74 & 221.9 & 4.7 & 224.665 & 2 & 9 & 1.927 & H 22.51 & \\
              &       &       &       & 294.7 & 5.1 & 294.523 & 1 & 6 & &&\\
              &       &       &       & 448.2 & 10.9 & 445.364 & 1 & 11 &&& \\
              &       &       &       & 789.5 & 10.2 & 791.684 & 1 & 21 &&& \\ 
              &       &       &       & 894.7 & 43.8 & 895.161 & 2 & 42 &&& \\
              &       &       &       & 968.7 & 6.3 & 968.218 & 2 & 46 &&& \\ 
\hline
 J2319$+$5153 & 11755 & 0.976 & -7.41  & 354.491 & 6.297 & 353.301 & 2 & 14 & 2.714 & H 6.18 & Castanheira et al. (2013b)\\
              &&&& 383.893 & 5.917 & 390.465 & 1 & 8 &&& \\
              &&&& 428.659 & 5.347 & 427.649 & 1 & 9 &&& \\
              &&&& 454.468 & 9.343 & 455.173 & 1 & 10 &&&\\
              &&&& 714.787 & 34.01 & 713.629 & 1 & 17 &&&\\ 
              &&&& 766.637 & 7.677 & 769.345 & 2 & 33 &&&\\             
              & 12104 & 1.024 & -5.70 & 354.491 & 6.297 & 355.722 & 1 & 8 & 2.576 & H 19.51 &\\
              &&&& 383.893 & 5.917 & 382.162 & 1 & 9 &&&\\
              &&&& 428.659 & 5.347 & 428.272 & 2 & 20 &&& \\
              &&&& 454.468 & 9.343 & 454.433 & 2 & 21 &&&\\
              &&&& 714.787 & 34.01 & 710.236 & 1 & 19 &&&\\ 
              &&&& 766.637 & 7.677 & 770.813 & 2 & 37 &&&\\
\hline
 J2350$-$0054 & 11082 & 1.024 & -7.43 & 273.3 & 6.2  & 272.840 & 2 & 10 & 0.644 & H 38.29 & Mukadam et al. (2004)\\
              &       &       &       & 304.3 & 17.0 & 304.115 & 1 & 5 & & &\\
              &       &       &       & 391.1 & 7.5  & 391.645 & 2 & 15 &&&\\
              & 11690 & 0.998 & -9.30 & 273.3 & 6.2  & 273.974 & 1 & 4 & 0.414 & SC 28.58 & \\
              &       &       &       & 304.3 & 17.0 & 303.793 & 1 & 5 && &\\
              &       &       &       & 391.1 & 7.5  & 391.135 & 2 & 14 &&&\\ 
              & 10225 & 0.660 & -7.33 & 273.3 & 6.2  & 272.951 & 2 & 6 & 0.628 && \\   
              &       &       &       & 304.3 & 17.0 & 305.474 & 1 & 4 &&&\\  
              &       &       &       & 391.1 & 7.5  & 390.196 & 1 & 5 &&& \\        
\hline
 J1916$+$3938 & 11391 & 0.837 & -7.36 & 823.9 & 0.44 & 825.248 & 1 & 18 & 1.522 & & Hermes et al. (2011)\\
              &       &       &       & 834.1 & 0.32 & 832.759 & 2 & 32 &       & & amplitudes are in \%\\
              &       &       &       & 934.5 & 0.36 & 932.693 & 2 & 36 &&&\\
              &       &       &       & 968.9 & 0.44 & 971.579 & 1 & 21 &&&\\
              &       &       &       & 1089.0 & 0.25 & 1089.111 & 1 & 24 &&&\\
              &       &       &       & 1169.9 & 0.23 & 1169.587 & 1 & 26 &&&\\
              &       &       &       & 1436.7 & 0.24 & 1437.089 & 2 & 56 &&&\\                          
\hline
 L19$-$2      & 12033 & 0.705 & -4.45 & 113.8 & 2.4 & 113.672 & 2 & 2 & 1.263 && Castanheira \& Kepler (2009)\\
              &       &       &       & 118.7 & 1.2 & 114.615 & 1 & 1 & &&\\
              &       &       &       & 143.6 & 0.6 & 143.499 & 2 & 3 & &&\\ 
              &       &       &       & 192.6 & 6.5 & 193.167 & 1 & 2 &&&\\
              & 11180 & 0.770 & -4.91 & 113.8 & 2.4 & 113.293 & 2 & 2 & 1.303 &&\\
              &       &       &       & 118.7 & 1.2 & 117.948 & 1 & 1 &&&\\ 
              &       &       &       & 143.6 & 0.6 & 145.601 & 2 & 3 &&&\\ 
              &       &       &       & 192.6 & 6.5 & 192.735 & 1 & 2 &&&\\
\hline
 G207$-$9     & 12030 & 0.837 & -6.36 & 259.1 & 17.3 & 258.853 & 1 & 4 & 1.025 & & Castanheira \& Kepler (2009)\\
              &       &       &       & 292.0 & 49.0 & 290.059 & 2 & 10 &&&\\
              &       &       &       & 318.0 & 64.0 & 318.256 & 1 & 5 &&&\\
              &       &       &       & 557.4 & 63.4 & 556.024 & 1 & 12 &&&\\
              &       &       &       & 740.4 & 46.4 & 740.499 & 1 & 17 &&&\\
\hline
EC0532$-$5605 & 11281 & 0.949 & -8.38 & 522.4 & 2.1 & 522.957 & 2 & 19 & 1.687 & SC 16.13 & Fontaine et al. (2003)\\
              &       &       &       & 563.7 & 2.5 & 565.264 & 2 & 21 &&&\\
              &       &       &       & 599.7 & 2.5 & 599.172 & 1 & 12 &&&\\
              &       &       &       & 686.1 & 5.5 & 681.622 & 1 & 14 &&&\\
              &       &       &       & 723.7 & 7.8 & 723.550 & 1 & 15 &&&\\
              &       &       &       & 753.8 & 4.8 & 753.922 & 2 & 28 &&&\\
              &       &       &       & 822.3 & 3.4 & 824.200 & 2 & 31 &&&\\
              &       &       &       & 881.7 & 2.9 & 881.062 & 2 & 33 &&&\\      
              & 11197 & 0.949 & -5.51 & 522.4 & 2.1 & 520.114 & 1 & 13 & 1.996 & H 2.02 & \\
              &       &       &       & 563.7 & 2.5 & 563.324 & 1 & 14 &&&\\
              &       &       &       & 599.7 & 2.5 & 610.122 & 1 & 15 &&&\\
              &       &       &       & 686.1 & 5.5 & 691.119 & 1 & 18 &&& \\
              &       &       &       & 723.7 & 7.8 & 721.031 & 1 & 19 &&&\\
              &       &       &       & 753.8 & 4.8 & 753.882 & 1 & 20 &&&\\
              &       &       &       & 822.3 & 3.4 & 824.317 & 2 & 38 &&&\\
              &       &       &       & 881.7 & 2.9 & 880.211 & 1 & 23 &&&\\   
\enddata
\label{tabla-periodos-1}
 \end{deluxetable}

\begin{deluxetable}{cccccccccccc}
\tabletypesize{\tiny}
\tablecaption {Same as Table \ref{tabla-periodos-1} but for the 24 massive ZZ Ceti stars with carbon$-$oxygen core, with less than three observed periods. }
\tablewidth{0pt}
\tablehead{
\colhead{${\rm star}$} & \colhead{$T_{\rm eff}$} & \colhead{$M_*/M_{\odot}$} & \colhead{$\log(M_{\rm H}/M_*)$} & \colhead{$\Pi_i^{\rm obs}$} &  \colhead{A}    & \colhead{$\Pi_k^{\rm th}$} & \colhead{$\ell$} & \colhead{$k$} & \colhead{$\Phi$} & \colhead{Crist.} & \colhead{Ref.}\\
    & $[$K$]$     &        &    &   $[$s$]$      & (mma) & $[$s$]$       &     &     &     & \% & }  
\startdata
 J0102$-$0032 & 10985 & 0.660 & -6.35 & 830.3 & 29.2 & 830.227 & 1 & 15 & 0.250 &  & Castanheira \& Kepler (2009)\\
              &       &       &       & 926.1 & 37.2 & 826.527 & 1 & 17 &&&\\
              & 11644 & 0.721 & -9.25 & 830.3 & 29.2 & 830.516 & 1 & 15 & 0.143 &&\\
              &       &       &       & 926.1 & 37.2 & 926.029 & 1 & 17 &&&\\
\hline
 J0111$+$0018 & 11826 & 0.800 & -8.34 & 255.50 & 12.95 & 254.883 & 1 & 3 & 1.737 && Mukadam et al. (2004)\\
              &       &       &       & 292.97 & 22.13 & 295.827 & 1 & 4 & &&\\
\hline
 J0249$-$0100 & 11177 & 0.632 & -4.86 & 1006.5 & 8.60 & 1006.741 & 1 & 20 & 0.129 && Castanheira \& Kepler (2009)\\
              &       &       &       & 1045.3 & 8.89 & 1045.316 & 1 & 21 &&&\\ 
              & 11015 & 0.660 & -6.35 & 1006.5 & 8.60 & 1006.268 & 1 & 19 & 0.261 &&\\
              &       &       &       & 1045.3 & 8.89 & 1045.134 & 2 & 35 & &&\\
              & 11310 & 0.837 & -9.34 & 1006.5 & 8.60 & 1006.662 & 1 & 21 & 0.239 && \\
              &       &       &       & 1045.3 & 8.89 & 1044.984 & 1 & 22 &&&\\ 
\hline
 J0303$-$0808 & 11178 & 0.917 & -7.39 &  707 &  4.1 & 706.454  & 1 & 15 & 0.462 & SC 5.47 & Castanheira et al. (2006)\\
              &&&                     &  1128 & 3.5 & 1128.220 & 2 & 44 & & & \\
              & 11755 & 0.917 & -6.43 &  707 &  4.1 & 706.667  & 2 & 21 & 0.502 & H (NCrist.) &\\
              &&&                     &  1128 & 3.5 & 1128.254 & 2 & 50 &&&\\
\hline
 J0322$-$0049 & 10967 & 0.660 & -7.33 & 767.5 & 15.1 & 767.449 & 1 & 13 & 0.051 && Mukadam et al. (2004)\\
              & 11016 & 0.770 & -4.91 & 767.5 & 15.1 & 767.573 & 1 & 17 & 0.073 &&\\
\hline
 J0349$+$1036 & 11724 & 0.878 & -7.38 & 184.50 & 3.76 & 184.508 & 1 & 1 & 0.009 & & Castanheira et al. (2013a)\\ 
\hline
 J0825$+$4119 & 11921 & 0.998 & -7.42 & 611.0 & 11.2 & 610.937 & 1 & 14 & 0.106 &  SC 26.20 & Mukadam et al. (2004)\\
              &       &       &       & 653.4 & 17.1 & 653.548 & 1 & 15 & & &  \\ 
              & 11750 & 0.998 & -9.31 & 611.0 & 11.2 & 610.994 & 1 & 13 & 0.210 & H 16.44  & \\
              &       &       &       & 653.4 & 17.1 & 653.764 & 1 & 14 & && \\
              & 11903 & 0.770 & -4.70 & 611.0 & 11.2 & 611.332 & 1 & 14 & 0.454 && \\
              &       &       &       & 653.4 & 17.1 & 653.976 & 1 & 15 & && \\
\hline
 J0843$+$0431 & 10995 & 0.837 & -5.00 & 1049 & 11.4 & 1048.458 & 1 & 26 & 0.279 & &Kepler et al. (2005)\\
              &       &       &       & 1085 & 7.42 & 1085.017 & 1 & 27 & &&\\
              & 10907 & 0.660 & -4.59 & 1049 & 11.4 & 1049.282 & 1 & 22 & 0.221 && \\
              &       &       &       & 1085 & 7.42 & 1084.905 & 2 & 40 & & &\\
              & 11417 & 0.770 & -9.33 & 1049 & 11.4 & 1048.507 & 2 & 36 & 0.500 && \\
              &       &       &       & 1085 & 7.42 & 1084.838 & 1 & 21 &&& \\         
\hline
 J0855$+$0635 & 10989 & 0.800 & -6.35 & 433 & 15 & 432.696 & 2 & 15 & 0.261 & & Castanheira et al. (2006)\\
              &       &       &       & 850 & 44 & 849.996 & 1 & 18 & &&\\
              & 11152 & 0.770 & -7.34 & 433 & 15 & 432.964 & 2 & 14 & 0.042 &&\\
              &       &       &       & 850 & 44 & 850.013 & 2 & 30 & &&\\ 
              & 11179 & 0.837 & -9.33 & 433 & 15 & 433.221 & 1 & 7 & 0.279 &&\\
              &       &       &       & 850 & 44 & 849.808 & 2 & 30 & &&\\ 
\hline
 J0923$+$0120 & 11123 & 0.949 & -9.29 & 595.055 & 2.73  & 595.960 & 1 & 12 & 0.172 &  H 6.46 & A. Mukadam, T. Metcalfe 2006\\
              &       &       &       & 1436.370 & 1.44 & 1436.112 & 1 & 31 & & &\\
              & 11007 & 0.800 & -6.35 & 595.055  & 2.73 & 595.206  & 1 & 12 & 0.118 && \\
              &       &       &       & 1436.370 & 1.44 & 1436.421 & 2 & 55 &&&\\
              & 11407 & 0.998 & -5.70 & 595.055  & 2.73 & 594.765  & 1 & 15 & 1.156 & SC 30.18 &\\
              &       &       &       & 1436.370 & 1.44 & 1436.383 & 2 & 66 &&&\\
\hline
 J0925$+$0509 & 10617 & 0.800 & -9.25 & 1159 & 2.72 & 1158.840 & 1 & 22 & 0.194 & & This paper \\
              &       &       &       & 1341 & 4.00 & 1341.229 & 1 & 26 & &&\\
              & 10930 & 0.800 & -8.34 & 1159 & 2.72 & 1158.874 & 1 & 23 & 0.332 &&\\
              &       &       &       & 1341 & 4.00 & 1341.357 & 1 & 27 &&&\\  
\hline
 J0939$+$5609 & 11672 & 0.770 & -4.91 & 249.9 & 7.2 & 249.897 & 1 & 4 & 0.003 & & Mukadam et al. (2004)\\
              & 11660 & 0.837 & -5.41 & 249.9 & 7.2 & 249.895 & 1 & 4 & 0.005 &&\\
              & 11339 & 0.705 & -4.88 & 249.9 & 7.2 & 249.901 & 1 & 3 & 0.001 &&\\
\hline
 J0940$+$0052 & 10817 & 0.770 & -4.91 & 254.98 & 17.13 & 255.118 & 1 & 4 & 0.002 & & Castanheira et al. (2013a)\\
              & 11091 & 0.721 & -6.43 & 254.98 & 17.13 & 255.118 & 1 & 3 & 0.003 &&\\
\hline
 J1105$-$1613 & 11336 & 0.837 & -7.36 & 192.66 & 12.09 & 192.581 & 1 & 2 & 0.632 & & Castanheira et al. (2010)\\
              &       &       &       & 298.25 & 7.09  & 298.223 & 2 & 10 &&&\\
              & 11811 & 0.721 & -9.25 & 192.66 & 12.09 & 192.528 & 1 & 1 & 0.187 && \\
              &       &       &       & 298.25 & 7.09  & 298.113 & 2 & 8 & &&\\
\hline
 J1218$+$0042 & 11321 & 0.705 & -7.35 & 258 & 16 & 258.000 & 1 & 3 & 0.0003 & & Kepler et al. (2005)\\
              & 10769 & 0.721 & -6.43 & 258 & 16 & 258.000 & 1 & 3 & 0.0007 &&\\
\hline
 J1337$+$0104 & 11245 & 0.837 & -8.34 & 715 & 10.0 & 715.001 & 1 & 14 & 0.001 & & Kepler et al. (2005)\\ 
              & 11622 & 0.878 & -5.07 & 715 & 10.0 & 715.002 & 2 & 33 & 0.004 & &\\ 
              & 11377 & 0.976 & -8.38 & 715 & 10.0 & 715.006 & 1 & 15 & 0.006 & SC 25.55 & \\
\hline
 J1612$+$0830 & 11818 & 0.705 & -4.45 & 115 & $\cdots$ & 114.982 & 1 & 1 & 0.018 & & Castanheira et al. (2013a)\\ 
\hline
 J1641$+$3521 & 11499 & 0.721 & -4.50 & 809.3 & 27.3 & 809.303 & 1 & 18 & 0.003 & & Castanheira et al. (2006) \\
              & 11367 & 0.878 & -5.40 & 809.3 & 27.3 & 809.297 & 1 & 21 & 0.003 & & \\
\hline
 J1650$+$3010 & 11169 & 1.024 & -5.74 & 339.06 & 14.71 & 339.065 & 1 & 7 & 0.005 &  SC 44.35 & Castanheira et al. (2006)\\ 
              & 11269 & 0.998 & -6.50 & 339.06 & 14.71 & 339.056 & 2 & 14 & 0.007 & H  20.97 &\\  
              & 11017 & 0.721 & -8.33 & 339.06 & 14.71 & 339.067 & 1 &  5  & 0.007 & &\\
\hline
 J2159$+$1322 & 11670 & 0.976 & -7.41 & 683.7 & 11.7 & 683.574 & 1 & 16 & 0.170 &  H 7.73 & Mullally et al. (2005)\\
              &       &       &       & 801.0 & 15.1 & 801.124 & 2 & 34 &       &         & \\
              & 11372 & 1.050 & -7.42 & 683.7 & 11.7 & 683.324 & 1 & 16 & 0.201 & SC 52.77 &\\
              &       &       &       & 801.0 & 15.1 & 800.974 & 1 & 19 &       & &\\    
\hline
 J2208$+$0654 & 11139 & 0.949 & -5.51 & 668.07 & 4.05 & 667.388 & 1 & 17 & 0.361 & H 2.86 & Castanheira et al. (2013a)\\
              &       &       &       & 757.23 & 4.46 & 757.190 & 1 & 20 &       & &\\
              & 11483 & 0.837 & -5.00 & 668.07 & 4.05 & 668.102 & 1 & 13 & 0.109 & &\\
              &       &       &       & 757.23 & 4.46 & 757.043 & 1 & 15 &       &&\\                  
              & 11058 & 0.917 & -9.28 & 668.07 & 4.05 & 668.799 & 2 & 13 & 0.338 & SC 7.41 &\\
              &       &       &       & 757.23 & 4.46 & 757.484 & 1 & 27 &       &       & \\
\hline
 J2214$-$0025 & 11635 & 0.878 & -7.38 & 195.2 & 6.1 & 195.365 & 1 & 2 & 0.102 & & Mullally et al. (2005) \\
              &       &       &       & 255.2 & 13.1 & 255.323 & 1 & 4 &&&\\      
              & 11358 & 0.837 & -8.37 & 195.2 & 6.1 & 195.137 & 2 & 5 & 0.096 &&\\
              &       &       &       & 255.2 & 13.1 & 255.282 & 1 & 3 & &&\\
\hline
 G2226$-$29   & 12270 & 0.770 & -4.69 & 109.278 & $\cdots$ & 109.246 & 1 & 1 & 0.032 & & Kepler et al. (2005)\\
\hline
 BPM 30551    & 11157 & 0.721 & -5.37 & 606.8 & 11.5 & 606.728 & 1 & 12 & 0.176  &  & Castanheira \& Kepler (2009)\\
              &       &       &       & 744.7 & 10.5 & 744.979 & 1 & 15 &        &&\\ 
              & 11436 & 0.705 & -5.36 & 606.8 & 11.5 & 607.055 & 1 & 12 & 0.175 && \\
              &       &       &       & 744.7 & 10.5 & 744.605 & 1 & 15 &        &&\\  
\enddata
\label{tabla-periodos-2}
 \end{deluxetable}

\begin{table*}
\caption{Structural parameters for the best fit models corresponding to each DAV  star analyzed in this paper. We list the star denomination, surface gravity, effective temperature, stellar mass, hydrogen and helium mass in terms of the stellar mass, luminosity, radius and oxygen central abundance by mass. The uncertainties are the internal errors of the fitting procedure.}
\scriptsize
\begin{tabular}{lcccccccc}
\hline\hline
     Star     &   $\log g$       &  $T_{\rm eff}$ [K]  &  $M_*/M_{\odot}$   & $M_{\rm H}/M_*$ &     $M_{\rm He}/M_*$   & $\log(L/L_{\odot})$ & $\log(R/R_{\odot})$ & $X_{\rm O}$\\
\hline\hline 
 J0000$-$0046 & $ 8.46\pm 0.03 $ & $11\,352\pm 53  $ & $0.878\pm 0.021$ & $ 5.17\times 10^{-10}$ & $2.59 \times 10^{-3} $ & $-2.909\pm 0.009$ & $-2.041\pm 0.014$ & $0.611$\\
 J0048$+$1521 & $ 8.57\pm 0.03 $ & $11\,470\pm 110 $ & $0.949\pm 0.014$ & $ 3.11\times 10^{-6}$ & $1.19 \times 10^{-3} $ & $-2.958\pm 0.017$ & $-2.075\pm 0.017$ & $0.614$\\
 J0102$-$0032 & $ 8.12\pm 0.03 $ & $10\,985\pm 43  $ & $0.660\pm 0.023$ & $ 4.46\times 10^{-7}$ & $1.22 \times 10^{-2} $ & $-2.747\pm 0.007$ & $-1.931\pm 0.011$ & $0.730$\\
 J0111$+$0018 & $ 8.34\pm 0.17 $ & $11\,826\pm 385 $ & $0.800\pm 0.019$ & $ 4.58\times 10^{-9}$ & $4.74 \times 10^{-3} $ & $-2.762\pm 0.056$ & $-2.001\pm 0.078$ & $0.648$\\
 J0249$-$0100 & $ 8.07\pm 0.02 $ & $11\,177\pm 34  $ & $0.632\pm 0.012$ & $ 1.40\times 10^{-5}$ & $1.75 \times 10^{-2} $ & $-2.678\pm 0.006$ & $-1.911\pm 0.009$ & $0.755$\\
 J0303$-$0808 & $ 8.52\pm 0.02 $ & $11\,178\pm 19  $ & $0.917\pm 0.020$ & $ 4.07\times 10^{-8}$ & $1.34 \times 10^{-3} $ & $-2.977\pm 0.003$ & $-2.061\pm 0.007$ & $0.609$\\
 J0322$-$0049 & $ 8.12\pm 0.06 $ & $10\,967\pm 58  $ & $0.660\pm 0.023$ & $ 4.70\times 10^{-8}$ & $1.22 \times 10^{-2} $ & $-2.754\pm 0.013$ & $-1.933\pm 0.023$ & $0.730$\\
 J0349$+$1036 & $ 8.46\pm 0.08 $ & $11\,724\pm 181 $ & $0.878\pm 0.021$ & $ 4.13\times 10^{-8}$ & $2.58 \times 10^{-3} $ & $-2.851\pm 0.027$ & $-2.040\pm 0.038$ & $0.611$\\
 J0825$+$0329 & $ 8.29\pm 0.03 $ & $11\,419\pm 57  $ & $0.770\pm 0.025$ & $ 4.28\times 10^{-6}$ & $5.97 \times 10^{-3} $ & $-2.781\pm 0.009$ & $-1.981\pm 0.012$ & $0.655$\\
 J0825$+$4119 & $ 8.65\pm 0.03 $ & $11\,921\pm 88  $ & $0.998\pm 0.013$ & $ 3.83\times 10^{-8}$ & $7.74 \times 10^{-4} $ & $-2.957\pm 0.013$ & $-2.107\pm 0.015$ & $0.629$\\
 J0843$+$0431 & $ 8.39\pm 0.04 $ & $10\,995\pm 47  $ & $0.837\pm 0.021$ & $ 9.96\times 10^{-6}$ & $3.18 \times 10^{-3} $ & $-2.913\pm 0.008$ & $-2.013\pm 0.018$ & $0.640$\\
 J0855$+$0635 & $ 8.34\pm 0.05 $ & $10\,989\pm 56  $ & $0.800\pm 0.019$ & $ 4.43\times 10^{-7}$ & $4.47 \times 10^{-3} $ & $-2.883\pm 0.009$ & $-1.999\pm 0.020$ & $0.648$\\
 J0923$+$0120 & $ 8.58\pm 0.02 $ & $11\,123\pm 30  $ & $0.949\pm 0.014$ & $ 5.18\times 10^{-10}$ & $1.20 \times 10^{-3} $ & $-3.023\pm 0.004$ & $-2.080\pm 0.010$ & $0.614$\\
 J0925$+$0509 & $ 8.34\pm 0.04 $ & $10\,617\pm 46  $ & $0.800\pm 0.019$ & $ 5.62\times 10^{-10}$ & $4.74 \times 10^{-3} $ & $-2.948\pm 0.008$ & $-2.002\pm 0.018$ & $0.648$\\
 J0939$+$5609 & $ 8.28\pm 0.12 $ & $11\,672\pm 239 $ & $0.770\pm 0.025$ & $ 1.23\times 10^{-5}$ & $5.96 \times 10^{-3} $ & $-2.738\pm 0.037$ & $-1.979\pm 0.053$ & $0.655$\\
 J0940$-$0052 & $ 8.29\pm 0.12 $ & $10\,817\pm 230 $ & $0.770\pm 0.025$ & $ 1.23\times 10^{-5}$ & $5.96 \times 10^{-3} $ & $-2.872\pm 0.037$ & $-1.980\pm 0.053$ & $0.655$\\
 J1105$-$1613 & $ 8.22\pm 0.14 $ & $11\,336\pm 133 $ & $0.837\pm 0.021$ & $ 4.32\times 10^{-7}$ & $3.19 \times 10^{-3} $ & $-2.757\pm 0.020$ & $-1.963\pm 0.036$ & $0.640$\\
 J1200$-$0251 & $ 8.52\pm 0.06 $ & $11\,715\pm 230 $ & $0.917\pm 0.020$ & $ 3.69\times 10^{-7}$ & $1.33 \times 10^{-3} $ & $-2.892\pm 0.034$ & $-2.059\pm 0.024$ & $0.609$\\
 J1216$+$0922 & $ 8.30\pm 0.10 $ & $11\,658\pm 205 $ & $0.770\pm 0.025$ & $ 4.57\times 10^{-8}$ & $5.97 \times 10^{-3} $ & $-2.753\pm 0.030$ & $-1.985\pm 0.044$ & $0.655$\\
 J1218$+$0042 & $ 8.20\pm 0.11 $ & $11\,321\pm 221 $ & $0.705\pm 0.023$ & $ 4.48\times 10^{-8}$ & $7.66 \times 10^{-3} $ & $-2.741\pm 0.035$ & $-1.954\pm 0.049$ & $0.661$\\
 J1222$-$0243 & $ 8.40\pm 0.06 $ & $11\,180\pm 85 $  & $0.837\pm 0.020$ & $ 4.41\times 10^{-8}$ & $3.19 \times 10^{-3} $ & $-2.893\pm 0.013$ & $-2.019\pm 0.024$ & $0.640$\\
 J1257$+$0124 & $ 8.18\pm 0.06 $ & $11\,172\pm 72  $ & $0.705\pm 0.023$ & $ 3.59\times 10^{-5}$ & $7.63 \times 10^{-3} $ & $-2.744\pm 0.018$ & $-1.944\pm 0.022$ & $0.661$\\
 J1323$+$0103 & $ 8.51\pm 0.04 $ & $11\,535\pm 72  $ & $0.917\pm 0.020$ & $ 3.90\times 10^{-6}$ & $1.31 \times 10^{-3} $ & $-2.914\pm 0.011$ & $-2.057\pm 0.016$ & $0.609$\\
 J1337$+$0104 & $ 8.40\pm 0.02 $ & $11\,245\pm 30  $ & $0.837\pm 0.021$ & $ 4.58\times 10^{-9}$ & $3.19 \times 10^{-3} $ & $-2.884\pm 0.004$ & $-2.020\pm 0.005$ & $0.640$\\
 J1612$+$0830 & $ 8.17\pm 0.19 $ & $11\,818\pm 350 $ & $0.705\pm 0.023$ & $ 3.59\times 10^{-5}$ & $7.63 \times 10^{-3} $ & $-2.645\pm 0.092$ & $-1.943\pm 0.087$ & $0.661$\\
 J1641$+$3521 & $ 8.20\pm 0.04 $ & $11\,499\pm 62  $ & $0.721\pm 0.025$ & $ 3.13\times 10^{-5}$ & $7.22 \times 10^{-3} $ & $-2.710\pm 0.010$ & $-1.952\pm 0.016$ & $0.659$\\
 J1650$+$3010 & $ 8.69\pm 0.05 $ & $11\,169\pm 100 $ & $1.024\pm 0.013$ & $ 1.83\times 10^{-6}$ & $5.56 \times 10^{-4} $ & $-3.099\pm 0.016$ & $-2.121\pm 0.026$ & $0.631$\\
 J1711$+$6541 & $ 8.61\pm 0.02 $ & $11\,280\pm 83  $ & $0.976\pm 0.014$ & $ 3.44\times 10^{-7}$ & $1.09\times 10^{-3} $ & $-3.0336\pm 0.013$ & $-2.721\pm 0.011$ & $0.613$\\
 J2128$-$0007 & $ 8.62\pm 0.10 $ & $11\,569\pm 150 $ & $0.976\pm 0.014$ & $ 5.12\times 10^{-10}$ & $1.10 \times 10^{-3} $ & $-2.985\pm 0.022$ & $-2.095\pm 0.051$ & $0.613$\\
 J2159$+$1322 & $ 8.57\pm 0.04 $ & $11\,670\pm 41  $ & $0.976\pm 0.014$ & $ 3.93\times 10^{-8}$ & $1.10 \times 10^{-3} $ & $-2.936\pm 0.007$ & $-2.079\pm 0.015$ & $0.613$\\
 J2208$+$0654 & $ 8.57\pm 0.04 $ & $11\,138\pm 61  $ & $0.949\pm 0.014$ & $ 3.11\times 10^{-6}$ & $1.19 \times 10^{-3} $ & $-3.012\pm 0.010$ & $-2.076\pm 0.019$ & $0.614$\\
 J2208$+$2059 & $ 8.74\pm 0.06 $ & $11\,355\pm 110 $ & $1.050\pm 0.013$ & $ 1.44\times 10^{-6}$ & $1.09 \times 10^{-3} $ & $-3.103\pm 0.017$ & $-2.138\pm 0.029$ & $0.613$\\
 J2209$-$0919 & $ 8.57\pm 0.04 $ & $11\,943\pm 102 $ & $0.949\pm 0.014$ & $ 3.66\times 10^{-8}$ & $1.19 \times 10^{-3} $ & $-2.897\pm 0.015$ & $-2.078\pm 0.018$ & $0.614$\\
 J2214$-$0025 & $ 8.40\pm 0.17 $ & $11\,635\pm 246 $ & $0.878\pm 0.021$ & $ 4.13\times 10^{-8}$ & $2.58 \times 10^{-3} $ & $-2.825\pm 0.038$ & $-2.019\pm 0.074$ & $0.611$\\
 J2319$+$5153 & $ 8.57\pm 0.06 $ & $11\,755\pm 98  $ & $0.976\pm 0.014$ & $ 3.93\times 10^{-8}$ & $1.10 \times 10^{-3} $ & $-2.925\pm 0.015$ & $-2.079\pm 0.023$ & $0.613$\\
 J2350$-$0054 & $ 8.70\pm 0.04 $ & $11\,082\pm 59  $ & $1.024\pm 0.013$ & $ 3.74\times 10^{-8}$ & $5.58 \times 10^{-4} $ & $-3.117\pm 0.009$ & $-2.123\pm 0.021$ & $0.631$\\
 J1916$+$3938 & $ 8.40\pm 0.03 $ & $11\,391\pm 50  $ & $0.837\pm 0.021$ & $ 4.41\times 10^{-8}$ & $3.16 \times 10^{-3} $ & $-2.860\pm 0.008$ & $-2.019\pm 0.010$ & $0.640$\\ 
 G226$-$29    & $ 8.28\pm 0.08 $ & $12\,270\pm 401 $ & $0.770\pm 0.025$ & $ 2.02\times 10^{-5}$ & $5.95 \times 10^{-3} $ & $-2.647\pm 0.056$ & $-1.977\pm 0.092$ & $0.655$\\
 L19$-$2      & $ 8.17\pm 0.08 $ & $12\,033\pm 316 $ & $0.705\pm 0.023$ & $ 3.59\times 10^{-5}$ & $7.63 \times 10^{-3} $ & $-2.613\pm 0.046$ & $-1.943\pm 0.037$ & $0.661$\\
 G207$-$9     & $ 8.40\pm 0.12 $ & $12\,030\pm 198 $ & $0.837\pm 0.021$ & $ 4.32\times 10^{-7}$ & $3.19 \times 10^{-3} $ & $-2.761\pm 0.029$ & $-2.017\pm 0.058$ & $0.640$\\
 EC0532$-$560 & $ 8.58\pm 0.05 $ & $11\,281\pm 64  $ & $0.949\pm 0.014$ & $ 4.16\times 10^{-9}$ & $1.20 \times 10^{-3} $ & $-2.998\pm 0.010$ & $-2.080\pm 0.022$ & $0.634$\\
 BPM30551     & $ 8.21\pm 0.07 $ & $11\,157\pm 106 $ & $0.721\pm 0.025$ & $ 4.31\times 10^{-6}$ & $7.25 \times 10^{-3} $ & $-2.772\pm 0.016$ & $-1.957\pm 0.029$ & $0.659$\\
\hline\hline 
\label{tabla-sismology}
\end{tabular}
\end{table*}


\end{document}